\documentclass[12pt]{article}
\usepackage{amsmath,amsfonts, amssymb, braket, bbold}
\usepackage{tikz}
\usetikzlibrary{trees,er,snakes,shapes,mindmap}
\usepackage{titlesec}
\titleformat{\section}
 {\normalfont\fontfamily{put}\fontsize{12pt}{16pt}\bfseries\color{black}}
{\thesection}{1em}{}
\titleformat{\subsection}
 {\normalfont\fontfamily{put}\fontsize{12pt}{16pt}\bfseries\color{black}}
{\thesubsection}{1em}{}
\def \beq  {\begin{equation}}
\def \eeq  {\end{equation}}
\def \beqar {\begin{eqnarray}}
\def \eeqar {\end{eqnarray}}
\allowdisplaybreaks
\def\sqr#1#2{{\vcenter{\vbox{\hrule height.#2pt
\hbox{\vrule width.#2pt height#1pt \kern#1pt
\vrule width.#2pt}\hrule height.#2pt}}}}

\def\la {{\langle}}
\def\ra {{\rangle}}

\def\vf {{\varphi}}

\def\Tr {{\rm Tr}}

\def\ba {\bar{a}}
\def\bD {\bar{D}}
\def\bA {\bar{A}}

\def\bx {\bar{x}}
\def\by {\bar{y}}

\def\bw {\bar{w}}

\def\bnabla {\bar{\nabla}}

\def\del {\partial}
\def\bdel{\bar{\partial}}
\def\a {\alpha}

\def\e {\epsilon}

\def\bz {{\bar{z}}}

\def\C {{\cal C}}
\def\D {{\cal D}}

\def\F {{\cal F}}
\def\G {{\cal G}}

\def\P {{\cal P}}

\def\vf {{\varphi}}

\def\half{\textstyle{1\over 2}}

\mathchardef\mhyphen="2D
\begin{document}
%%%%%%%%%%%%%%%%%%%%%%%%%%%%%%%%%%%%%%%%%%%%%%%
%%%%%%%%%%%%%%%%%%%%%%%%%%%%%%%%%%%%%%%%%%%%%%%
\fontfamily{bch}\fontsize{12pt}{16pt}\selectfont
%\fontfamily{pnb}\fontsize{12pt}{16pt}\selectfont
%\fontfamily{pzc}\fontsize{14pt}{16pt}\selectfont
%\fontfamily{pbk}\fontsize{12pt}{16pt}\selectfont
%\fontfamily{cmr}\fontsize{11pt}{15pt}\selectfont
%\fontfamily{put}\fontsize{12pt}{17pt}\selectfont
%\fontfamily{lmss}\fontsize{11pt}{16pt}\selectfont
%\fontfamily{phv}\fontshape{ro}\fontsize{11pt}{14pt}\selectfont
%\fontfamily{ptm}\fontseries{m}\fontshape{r}\fontsize{12pt}{16pt}\selectfont
%\fontfamily{pnc}\fontseries{m}\fontshape{r}\fontsize{11pt}{15pt}\selectfont
%\fontfamily{ppl}\fontseries{m}\fontshape{r}\fontsize{11pt}{15pt}\selectfont
%\usefont{T1}{phv}{m}{it}
%%%%%%%%%%%%%%%%%%%%%%%%%%%%%%%%%%%%%%%%%%%%%%%
%%%%%%%%%%%%%%%%%%%%%%%%%%%%%%%%%%%%%%%%%%%%%%%
\def \CMP {{Commun. Math. Phys.}}
\def \PRL {{Phys. Rev. Lett.}}
\def \PL {{Phys. Lett.}}
\def \NPBProc {{Nucl. Phys. B (Proc. Suppl.)}}
\def \NP {{Nucl. Phys.}}
\def \RMP {{Rev. Mod. Phys.}}
\def \JGP {{J. Geom. Phys.}}
\def \CQG {{Class. Quant. Grav.}}
\def \MPL {{Mod. Phys. Lett.}}
\def \IJMP {{ Int. J. Mod. Phys.}}
\def \JHEP {{JHEP}}
\def \PR {{Phys. Rev.}}
\def \JMP {{J. Math. Phys.}}
\def \GRG{{Gen. Rel. Grav.}}
%%%%%%%%%%%%%%%%%%%%%%%%%%%%%%%%%%%%%%%%%%%%%%%
%%%%%%%%%%%%%%%%%%%%%%%%%%%%%%%%%%%%%%%%%%%%%%%
\begin{titlepage}
\null\vspace{-62pt} \pagestyle{empty}
\begin{center}
%\rightline{CCNY-HEP-18/4}
%\rightline{August 2018}
\vspace{1.3truein} {\large\bfseries
Gauge and Scalar Fields on $\mathbb{CP}^2$: A Gauge-invariant Analysis}
\vskip .2in
{\large\bfseries I. The effective action from chiral scalars}\\
{\Large\bfseries ~}\\
%%%%%%%%%%%%%%%%%%%%%%%%%%%%%%%%%%%%%%%%%%%%%%%
%%%%%%%%%%%%%%%%%%%%%%%%%%%%%%%%%%%%%%%%%%%%%%%
{\sc Dimitra Karabali$^{a,c}$, Antonina Maj$^{a,b,c}$, V.P. Nair$^{b, c}$}\\
\vskip .2in
{\sl $^a$Physics and Astronomy Department,
Lehman College, CUNY\\
Bronx, NY 10468}\\
\vskip.1in
{\sl $^b$Physics Department,
City College of New York, CUNY\\
New York, NY 10031}\\
\vskip.1in
{\sl $^c$The Graduate Center, CUNY\\
New York, NY 10016}\\
 \vskip .1in
\begin{tabular}{r l}
{\sl E-mail}:&\!\!\!{\fontfamily{cmtt}\fontsize{11pt}{15pt}\selectfont 
dimitra.karabali@lehman.cuny.edu}\\
&\!\!\!{\fontfamily{cmtt}\fontsize{11pt}{15pt}\selectfont amaj@gradcenter.cuny.edu}\\
&\!\!\!{\fontfamily{cmtt}\fontsize{11pt}{15pt}\selectfont vpnair@ccny.cuny.edu}\\
\end{tabular}
\vskip .5in

%%%%%%%%%%%%%%%%%%%%%%%%%%%%%%%%%%%%%%%%%%%%%%%
%%%%%%%%%%%%%%%%%%%%%%%%%%%%%%%%%%%%%%%%%%%%%%%
\centerline{\large\bf Abstract}
\end{center}
A parametrization of gauge fields on complex projective spaces of
arbitrary dimension is given as a generalization of
 the real two-dimensional case. Gauge transformations act homogeneously on the fields, facilitating a manifestly gauge-invariant analysis.
 Specializing to four dimensions, we consider the nature of the effective action due to chiral scalars interacting with the gauge fields. The key qualitatively significant terms include a possible
 gauge-invariant mass term and a finite four-dimensional
 Wess-Zumino-Witten (WZW) action. We comment on relating the mass term to lattice simulations as well as Schwinger-Dyson analyses, and also on relating the WZW action to the instanton liquid picture of QCD.

\end{titlepage}
%%%%%%%%%%%%%%%%%%%%%%%%%%%%%%%%%%%%%%%%%%%%%%%
%%%%%%%%%%%%%%%%%%%%%%%%%%%%%%%%%%%%%%%%%%%%%%%
\fontfamily{bch}\fontsize{12pt}{17pt}\selectfont
\pagestyle{plain} \setcounter{page}{2}
\section{Introduction}

The gauge-invariant analyses of the low energy or long distance properties of
nonabelian gauge theories remains a challenging problem even after decades of work. Large scale numerical simulations have produced
important insights as well as quantitative estimates of physically relevant
observables, but the analytic understanding of the problem is far from satisfactory.
Perhaps the most revelatory aspect of this state of affairs is 
concerning the foundational ingredient needed for the
quantum theory, namely, the volume element for
the gauge-orbit space.
This space
is the set of all gauge potentials ($ {\cal A}$) modulo the set of all gauge transformations 
which are fixed to be identity at one point on the spacetime manifold (${\cal G}_*$) \cite{singer}.
Thus it is this space of gauge-invariant field configurations
over which the functional integration for such theories has to be carried out
to define the quantum theory;
i.e., the volume element of this gauge-orbit space 
$\C = {\cal A }/ {\cal G}_*$ provides the measure of integration.
There is still no satisfactory and explicit formula for this in
the continuum four-dimensional theory.
One can use gauge-fixing and the Faddeev-Popov procedure to construct this volume element
for a local section of ${\cal A}$ viewed as a ${\cal G}_*$-bundle over 
$\C$, or, equivalently, one may use the BRST (Becchi-Rouet-Stora-Tyutin) procedure. 
However, nonperturbative questions are generally beyond the reach of
this procedure, although
it may be adequate for the perturbative calculations.

In contrast to this, for gauge fields in two dimensions, the volume element for $\C$ can be calculated exactly in terms of a Wess-Zumino-Witten (WZW) action \cite{gawe}. 
Although there are no propagating degrees of freedom for gauge fields
in two dimensions, the result is relevant for
the Chern-Simons-WZW relationship \cite{witten} and in the solution of Yang-Mills theory on Riemann surfaces \cite{2dYM}.
This result may be taken as applying to the fields on a spatial slice in
(2+1)-dimensions, and one can thus seek to utilize it in a Hamiltonian
approach to  (2+1)-dimensional Yang-Mills theories.
Such an analysis has led to a
formula for the string tension and also provided insights
into the mass gap \cite{{KKN}, {nair-trento1}}, including supersymmetric cases \cite{AN1}.
The expression for the string tension agrees very well with
estimates via lattice simulations \cite{teper} and, more recently, estimates of the Casimir energy have provided independent verification of the mass gap (or the mass
defined by the propagator) \cite{chernodub}.

The calculation of the volume element of $\C$ in two dimensions
was made possible by a parametrization of the gauge fields 
which relied on the fact that
the two-dimensional space could be considered as a complex manifold.
For ${\mathbb{R}}^4$, there is no unique
complex structure, since there are many ways to pair the coordinates to form
complex ones. One could consider a twistor space version which would include the set  of all local complex structures. Calculations for the gauge theory would also require an infrared cutoff, so a compact space of
finite volume is a better alternative. The simplest case of such a space would
be the complex projective plane ${\mathbb{CP}}^2$, which is a complex K\"ahler manifold. The standard metric for this space is the Fubini-Study metric which is given
in local coordinates $z^a$, $\bz^{\bar a}$, $a= 1,2$, $\ba = 1, 2$, as
\beq
ds^2 = {dz \cdot d\bz \over (1+ z\cdot \bz/r^2)}
- {\bz \cdot dz \, z\cdot d\bz \over r^2 (1+ z\cdot \bz /r^2)^2}
= g_{a {\bar a}} dz^a d\bz^{\bar a}
\label{0a}
\eeq
where we have also included a scale parameter $r$ for the coordinates. 
The volume of $\mathbb{CP}^2$ with this metric is
$\pi^2 r^4 /2$, so $r$ can serve as an infrared cutoff.
As $r \rightarrow \infty$, the metric becomes that of flat space (although there are some global issues which will not be important for us).
Thus this space has a complex structure (which can help with
the parametrization of the fields) and a finite volume, with
a well-defined limit to the flat case. Indeed, a parametrization of the gauge 
potentials, as a generalization of the parametrization in
two dimensions, was given in \cite{nair}, where some 
preliminary results regarding the volume element
of $\C$ were also given.
Admittedly, the group of isometries
for the space is $SU(3)$, rather than the 4d Euclidean group.
However, this should not be an issue for many questions of interest.
Recall that one can obtain insights into the physics by studying lattice gauge theories even though the lattice breaks the Euclidean invariance, recovering
it only in the continuum limit. The analogue for our use of
$\mathbb{CP}^2$
would be the large $r$ limit. Also, there are many other instances in which
scenarios with reduced isometries can give insight into the physics of 
a problem, the Casimir effect being a classic example.

A closely related issue is the nature of quantum corrections
(to the gauge field dynamics) due to matter fields.
The calculation of such corrections in a manifestly gauge-invariant way 
using the parametrization mentioned above, can give insights into the renormalization structure of the gauge theory and hence to
some questions of physical interest.
We propose to take up a more detailed analysis of the volume
element for $\C$ and the nature of quantum corrections due to a
scalar field on $\mathbb{CP}^2$.
The present article will be devoted to the general framework and
the corrections due to the scalar field, with more details
on the volume element for $\C$ to be given in a follow-up paper \cite{KMN2}.

There are two physical aspects of nonabelian gauge theories
for which our analysis can lead to useful insights.
The first is about a possible mass term for gluons.
There has been growing evidence, based on 
analytical and numerical studies,
that the gluon acquires a ``mass"
\cite{{cornwall},{CPB},{ABP}}.
Given these results, one can ask if we can find
any evidence for a mass term in
a manifestly gauge-invariant and analytic approach.
This is what our analysis addresses.
One of the terms we find is indeed a mass term consistent with 
gauge invariance and all the isometries of the underlying space
$\mathbb{CP}^2$. 

The second consideration is about the instanton liquid picture
\cite{{schafer},{athenodorou}}.
Analytical investigations as well as lattice simulations have shown that
the infrared behavior of correlation
functions for gluons, and for hadrons, is dominated by a dense
collection of instantons, the so-called instanton liquid.
Again, in a manifestly gauge-invariant analysis,
one can ask if there are indications of an instanton liquid.
Indeed, we find that one of the terms in the effective action
is a four-dimensional Wess-Zumino-Witten (WZW) action
whose critical points are antiself-dual instantons \cite{{Don},{NS}}.
We will comment on these issues in more detail later.

The organization of this paper is as follows.
In section 2, we give the general parametrization of the fields,
discussing in turn the nature of scalars, vectors and gauge potentials
on $\mathbb{CP}^2$. We also introduce the action for the scalar field.
Section 3 is devoted to the calculation of the quantum corrections due
to the scalar field.
After giving the general framework,  we calculate the scalar field propagator
for $\mathbb{CP}^2$ and discuss regularization issues.
The leading terms among the quantum corrections are then obtained.
These include a WZW action with a finite coefficient, a quadratically 
divergent mass term, and
the expected log-divergence of the wave function renormalization for
the gauge fields.
In section 4, we discuss the physical implications.
Several computational details are given in appendixes, three of them,
so as to avoid clutter and keep an uninterrupted
flow of the general arguments in the text. Appendix A gives the 
parametrization of the gauge fields and the calculation of the scalar propagator on $\mathbb{CP}^k$, for arbitrary $k$, even though
$k =2$ is what is used in the text of the paper.
In Appendix B, we give the calculation of a current relevant for the
identification of the WZW term. Some of the subtleties regarding the WZW term are discussed in detail.
The ultraviolet divergences are calculated in Appendix C.

\section{Parametrization of fields}

As stated in the introduction, the manifold $\mathbb{CP}^2$ allows for a
parametrization of the fields with a clear separation of the
gauge-invariant degrees of freedom. This is most conveniently done in
terms of the coset structure of the space as
$\mathbb{CP}^2 = SU(3)/ U(2)$.
The manifold may be coordinatized in terms of a group
element $g \in SU(3)$, with $U(2) \subset SU(3)$ as the
local isotropy group and the coset directions corresponding to the
translational
degrees of freedom.
This shows that functions on $\mathbb{CP}^2$ correspond to functions on
$SU(3)$ which are invariant under $U(2)$, while vectors, tensors, etc. 
transform as
specific nontrivial representations of $U(2)$.

Turning to more specific details,
the defining fundamental representation is taken as a
$3\times 3$ unitary matrix $g$ 
 of unit determinant. It
can be parametrized as $g = \exp ({i t_a \, \vf^a})$, where $t_a$ form a basis for traceless hermitian
$3\times 3$ matrices, with $\Tr\,( t_a t_b) = \half \, \delta_{ab}$, and $\vf^a$ are the coordinates
 for $SU(3)$.
Following the familiar nomenclature from the
quark model,
we shall refer to the $SU(2)$ part of the $U(2)$ subgroup
as isospin (denoted by $I$) and
the $U(1)$ part of $U(2)$ as hypercharge (denoted by $Y$).
The subgroup $SU(2)$ corresponds to the directions
$a= 1, 2, 3$,
with the generators $t_1, \, t_2,\, t_3$ for its Lie algebra;
the hypercharge
corresponds to $2\,t_8/\sqrt{3}$.
On $g$, we can define left ($L_a$) and right ($R_a$)
translation operators by
\beq
L_a \, g = t_a \, g, \hskip .3in
R_a\, g = g \, t_a
\label{cp2-1}
\eeq
The translation operators (or derivative operators)
on $\mathbb{CP}^2$ can then be
defined as
\beq
R_{\pm 1} = R_4 \pm i R_5, \hskip .3in
R_{\pm 2} = R_6 \pm i R_7
\label{cp2-2}
\eeq
These are the appropriate complex components, 
we shall denote them by $R_i$, $R_{\bar i}$, $i, {\bar i} =
1, 2$. The matrices corresponding to these combinations
have all elements equal to zero, except for the $(i 3)$ 
and $(3 i)$ elements, which are equal to 1 for 
$R_i$ and $R_{\bar i}$, respectively.
The curvatures for $\mathbb{CP}^2$ take
values in the Lie algebra of $U(2)$, with the
operators $R_\a$, $\a= 1, 2, 3$, and $R_8$ defining the
analogue of spin.
Explicitly, $R_a$ can be realized as differential operators,
\beq
g^{-1} \, dg = - i t_a \, E^a_i \, d\vf^i,
\hskip .2in R_a = i (E^{-1})^i_a \, {\del \over \del \vf^i}
\label{cp2-3}
\eeq

A basis for functions on $SU(3)$ is given by
the finite-dimensional unitary representation matrices for $SU(3)$, 
denoted by
$\D_{AB}^{(s)}(g)= \la s, A \vert {\hat g} \vert s, B\ra$ (and often referred to as
the Wigner functions). The action of $R_a$ on these functions is given by
\beq
R_a \,\la s, A \vert {\hat g} \vert s, B\ra
= \la s, A \vert {\hat g} \, T_a \vert s, B\ra
= \la s, A \vert {\hat g} \vert s, C\ra\, (T_a)_{CB}
\label{cp2-4}
\eeq
where $T_a$ are the matrix representatives of $t_a$ in the
representation labeled by $s$. Functions, vectors and tensors
on $\mathbb{CP}^2$ have the mode expansion
\beq
F (g) = \sum_{s, A} C^{(s)}_{A} \, D^{(s)}_{A, w}(g)
= \sum_{s, A} C^{(s)}_{A} \, \la s, A \vert {\hat g} \vert s, w\ra
\label{cp2-5}
\eeq
where the states on the right, namely,
$\vert s, w\ra$ must be so chosen as to give the correct transformation property for $F(g)$ under $U(2) \in SU(3)$.

\subsection{Functions on $\mathbb{CP}^2$}

Functions on $\mathbb{CP}^2$ must be invariant under $U(2)$, so
we need states $\vert s, w\ra$ with
$Y =0$ and $I =0$. 
 A state $\vert \{a_i\}, \{ b_i\}\ra$ which carries a general $SU(3)$ representation is of the form
$T^{a_1 a_2 \cdots a_p}_{b_1 b_2 \cdots b_q}$, $a_i,\, b_j = 1,2, 3$, which we refer to as
a $(p,q)$-type representation. These are totally symmetric in all the upper indices
$a_i$'s and totally symmetric in all
the lower indices $b_j$'s with the trace (or any contraction between any choice of upper and lower indices) vanishing.
The $SU(3)$ action on $T^{a_1 a_2 \cdots a_p}_{b_1 b_2 \cdots b_q}$
is given by
\beq
T^{a_1 a_2 \cdots a_p}_{b_1 b_2 \cdots b_q}
\rightarrow  \bigl( g^{*a_1 a'_1}\,g^{*a_2 a'_2}\cdots \bigr)
\bigl( g_{b_1 b'_1}\,g_{b_2 b'_2}\cdots \bigr)
T^{a'_1 a'_2 \cdots a'_p}_{b'_1 b'_2 \cdots b'_q}
\label{cp2-6}
\eeq
Notice that the isospin subgroup acts on
indices taking values $1, 2$, while the
value of hypercharge is given as
\beq
Y=\begin{cases} - {1\over 3} \hskip .05in &a_i = 1, 2\\
~~~{2\over 3} & a_i =3
\end{cases},
\hskip .3in
Y=\begin{cases}  ~~{1\over 3} \hskip .05in& b_i = 1, 2\\
-{2\over 3} & b_i =3
\end{cases}
\label{cp2-6a}
\eeq
The choice of all indices equal to $3$ with $p =q$ corresponds to
the $U(2)$ invariant choice. Thus, for functions
on $\mathbb{CP}^2$, we need representations of the $(p,p)$-type
with the
mode expansion
given by
\beqar
f (g) 
&=& \sum_{s, A} C^{(p,p)}_{A} \, \la s, A \vert {\hat g} \vert 0 \ra
\nonumber\\
\vert 0\ra &\equiv& 
\vert (p,p), w\ra = \vert  3 3 3 \cdots,  3 3 3 \cdots \ra
\label{cp2-7}
\eeqar
For brevity, we will denote the $U(2)$-invariant state as $\ket{0}$.

\subsection{Vectors on $\mathbb{CP}^2$}

The translation operators $R_{+i}= R_i$ and $R_{-i}= R_{\bar i}$ transform as doublets of
$SU(2)$ and carry hypercharge $Y = 1, -1$, respectively.
Thus vectors on $\mathbb{CP}^2$ must have a similar transformation property.
This can be obtained for representations of the
$(p,p)$-type with $\vert s, w\ra$ of the form
$\vert 3 3 \cdots ,  i 3 3 \cdots \ra$ and
$\vert  i 3 3 \cdots, 3 3 \cdots \ra$, corresponding to
$Y= 1$ and $Y= -1$, respectively.
These can be obtained from the invariant state
$\vert 3 3 \cdots,   3 3 \cdots \ra$ by the application of 
$R_i$ and $R_{\bar i} $ respectively. The corresponding vectors
are the gradients of functions on $\mathbb{CP}^2$.

One can also obtain the required states from representations
of the
$(p+3,p)$-type with $\vert s, w\ra = \vert i 3 3 \cdots,  3 3 \cdots\ra$,
with $i = 1, 2$, corresponding to $Y = 1$,
and from $(p, p+3)$-type with
$\vert s, w\ra = \vert 3 3 \cdots, \,  i 3 3 \cdots\ra$
with $Y = -1$. 
Thus a vector on $\mathbb{CP}^2$ may be parametrized as
\beqar
A_i &=& - R_i \, f - \eta_{i {\bar i}} \e^{{\bar i}{\bar j}} \sum_{s,A} C_A^{(s)} \,\la s, A\vert \,{\hat g}\,  \vert  
{\bar j}  3 3 \cdots,  3 3 \cdots\ra\nonumber\\
\bA_{\bar i}  &=& - R_{\bar i} \, {\bar f} - \eta_{{\bar i} i} \e^{ij} \sum_{s^*,A} C_A^{(s^*)} \,\la s^*, A\vert \,{\hat g} \,\vert 
3 3 \cdots, j 3 3 \cdots\ra
\label{cp2-8}
\eeqar
where, on the right hand side, $s$ indicates representations
of the $(p+3, p)$-type and $s^*$ indicates
the $(p, p+3)$-type.
The first terms on the righthand side correspond to gradients of 
a function.
The $(p+3, p)$-type state $\vert  {\bar j}  3 3 \cdots,  3 3 \cdots\ra$ can be obtained from the
$SU(2)$ invariant states, with all indices equal to 3,
by the application of $R_{\bar j}$ operators.\footnote{As defined
in (\ref{cp2-6}), the state $\vert  j 3 3 \cdots,  3 3 \cdots\ra$ transforms under $SU(2)$ as the conjugate of the standard doublet representation, (emphasized by using
${\bar j}$), the
extra factor of $\eta_{i {\bar i}} \e^{{\bar i} {\bar j}}$ converts the transformation to the usual doublet
form.}
Specifically,
we can write
\beq
\eta_{i {\bar i}} \e^{{\bar i} {\bar j}} \vert  {\bar j}  3 3 \cdots,  3 3 \cdots\ra
= \eta_{i {\bar i}} \,\epsilon^{{\bar i} {\bar j} } \,R_{\bar j} \,\vert
3 3 \cdots, 3 3\cdots\ra
\label{cp2-9}
\eeq
where $\eta_{i{\bar i}} = \delta_{i {\bar i}}$ (which is the metric
for $\mathbb{CP}^2$ in the tangent frame) and
$\epsilon^{{\bar i} {\bar j} }$ is the Levi-Civita tensor.
The $SU(2)$ invariant state on the righthand side has
$Y = 2 $, so the corresponding term
in (\ref{cp2-8}) may be written as $\eta_{i{\bar i}} \epsilon^{{\bar i} {\bar j}}
R_{\bar j} \chi$ where $\chi$ has $Y = 2$.
(We may regard $ \epsilon^{{\bar i} {\bar j}} \chi$ as a rank-2 tensor
of the antiholomorphic type, so that the relevant term
in (\ref{cp2-8}) is the divergence
of an antisymmetric tensor.)
Similar statements hold for conjugates in the second line of
(\ref{cp2-8}), so that we can write the general parametrization as
\beqar
A_i &=& -R_i \, f - \eta_{i{\bar i} } \epsilon^{{\bar i} {\bar j}} \, R_{\bar j} \chi
\nonumber\\
\bA_{\bar i} &=& -R_{\bar i} \, {\bar f}  - \eta_{{\bar i}i } \epsilon^{i j} \, R_{ j} {\bar\chi}
\label{cp2-10}
\eeqar
These can be written in terms of the standard covariant derivatives
on $\mathbb{CP}^2$. $R_i$, $R_{\bar i}$ correspond to
the tangent frame, with
\beq
R_i = (e^{-1})^m_i \, \nabla_m , \hskip .3in
R_{\bar i} = - (e^{-1})^{\bar m}_{\bar i} \, \bnabla_{\bar m}
\label{cp2-11}
\eeq
Here $\nabla$'s include the spin connection as needed for 
$\chi$ and $\bar \chi$.
$e^i_m$ is the frame
field for the metric on $\mathbb{CP}^2$, i.e., $\eta_{i {\bar i}} e^i_m e^{\bar i}_{\bar m}
= g_{m {\bar m}}$.
The explicit formulae for the frame field and its inverse
for the Fubini-Study metric
(\ref{0a}) are
\beqar
e^a_m &=& {\delta^a_m \over \sqrt{1+ \bz\cdot z}}
- {\eta_{m {\bar m}} \bz^{\bar m} z^a \over (1 + \bz \cdot z) 
(1 + \sqrt{1 + \bz \cdot z})}\nonumber\\
(e^{-1})^m_a &=& \sqrt{1 + \bz \cdot z} \left[ \delta^m_a + 
{\eta_{a {\bar a}} \bz^{\bar a} z^m \over  1 + \sqrt{1+ \bz \cdot z} }
\right]
\label{cp2-11a}
\eeqar
In a coordinate basis, the parametrization (\ref{cp2-10}) takes the form
\beq
A_k = -\nabla_k \,f + g_{k {\bar k}} \e^{{\bar k} \bar m} \bnabla_{\bar m}
\chi, \hskip .3in
\bA_{\bar k}  = \bnabla_{\bar k}  \,{\bar f} - g_{{\bar k}k} \e^{k m} \nabla_{m}
{\bar \chi}
\label{cp2-12}
\eeq
In this paper,
we choose the components of $A$ to be related by
$(A_i)^\dagger = - \bA_{\bar i}$. This is in conformity with 
the use of antihermitian
components for the gauge fields, which is what is conventionally done for
nonabelian fields.

If we scale $z \rightarrow z/r$ and consider large values of $r$,
$\mathbb{CP}^2$ reduces to a flat space, but with a complex structure
since we still retain complex combinations of the real coordinates.
In this case, (\ref{cp2-12}) still retains its form,
\beq
A_k = -\del_k \,f + \eta_{k {\bar k}} \e^{{\bar k} \bar m} \bdel_{\bar m}
\chi, \hskip .3in
\bA_{\bar k}  = \bdel_{\bar k}  \,{\bar f} - \eta_{{\bar k}k} \e^{k m} \del_{m}
{\bar \chi}
\label{cp2-12a}
\eeq
An {\it a priori} and direct
demonstration that
this provides a complete and unique (see later) parametrization
of the fields in the flat space limit is difficult without the group theoretic arguments
which were used for $\mathbb{CP}^2$.

\subsection{Gauge fields on $\mathbb{CP}^2$}

We can use (\ref{cp2-12}) as the parametrization for Abelian gauge fields
(vector potentials) on $\mathbb{CP}^2$.
In this case, $\chi$, ${\bar \chi}$ and the real part of 
$f$ correspond to gauge-invariant degrees of freedom, while the
imaginary part of $f$ is the gauge parameter.

In generalizing to the nonabelian case, we first note that
the product of two functions on $\mathbb{CP}^2$ is still a function since it
remains invariant under $U(2)$. So we can compose functions. Likewise
the product of functions with $\chi$ or $\bar \chi$ retain the
same $U(2)$ transformations as $\chi$ and $\bar \chi$.
We can now write the generalization of (\ref{cp2-12})
to the nonabelian gauge
fields as
\beqar
A_i = - \nabla_i M \, M^{-1} + g_{i {\bar i}}  {\bar D}_{\bar j} \phi^{{\bar i} {\bar j}}\nonumber\\
\bA_{\bar i} = M^{\dagger -1} \bnabla_{\bar i} M^\dagger - g_{{\bar i} i}
D_j { \phi}^{\dagger ij}
\label{cp2-13}
\eeqar
Here $M$ and $M^\dagger$ are complex matrices
which are group elements in the complexification of the gauge group.
We will take the gauge group to be $SU(N)$ for simplicity.
(This is easily generalized to any Lie group.) In this case, 
$M$ and $M^\dagger$ are complex $N\times N$ matrices
which may be viewed as elements
of $SL(N, \mathbb{C})$.
Further, $\phi^{{\bar i} {\bar j} } = \epsilon^{{\bar i} {\bar j} } \phi$,
$\phi^{\dagger i j} = \e^{i j} \phi^\dagger$, where
$\phi^{{\bar i}{\bar j}}$, $\phi^{\dagger i j}$ are tensors valued in the Lie algebra of the gauge group $SU(N)$, in agreement with
$A_i$ and $\bA_{\bar i}$ being Lie algebra valued.
Since $\phi$ is complex, we may also view it as an element of the Lie algebra of
$SL(N, \mathbb{C})$, with $\phi^\dagger$ as its conjugate.
The derivatives $D_j$ and $\bD_{\bar j}$ are defined
by
\beq
D_j \Phi = \nabla_j \Phi + [-\nabla_j M M^{-1}, \Phi],\hskip .3in
\bD_{\bar j} \Phi = \bnabla_{\bar j} \Phi + [
M^{\dagger -1} \bnabla_{\bar j} M^\dagger, \Phi]
\label{cp2-14}
\eeq
acting on a field $\Phi$ which transforms under the adjoint representation
of the gauge group, $\Phi \rightarrow U \, \Phi \, U^\dagger$,
where $U\in SU(N)$ is the gauge transformation.
The potentials in (\ref{cp2-13}) transform as connections
with $M \rightarrow U\, M$, $M^\dagger \rightarrow M^\dagger U^\dagger$,
$(\phi, {\phi^\dagger} ) \rightarrow U (\phi, {\phi^\dagger} )  U^\dagger$.
The use of just $- \nabla_j M \, M^{-1}$,  $M^{\dagger -1} \bnabla_{\bar j} M^\dagger$
in defining $D_j$ and $\bD_{\bar j}$ suffices to ensure
that $D_j \Phi$ and $\bD_{\bar j} \Phi$ transform
covariantly under gauge transformations.\footnote{An important point about gauge fields is that the space of connections 
is an affine space, so that
one can reach
any point in this space from any other point by a straight line.
In other words, if
$A^{(1)}$ and $A^{(2)}$ denote two potentials, then
$A^{(1)} \, \tau + A^{(2)} \, (1-\tau )$, $0\leq \tau\leq 1$ transforms like a connection
for all $\tau$. Therefore one
can use a specific connection as a starting point and obtain every
other connection by adding something that transforms covariantly.
We may view $(-\nabla_i M M^{-1}, M^{\dagger -1} \bnabla_{\bar i} M^\dagger )$ as the starting connection and 
$(g_{i {\bar i}}  {\bar D}_{\bar j} \phi^{{\bar i} {\bar j}}, 
- g_{{\bar i} i}
D_j { \phi}^{\dagger ij} )$ as what is added. In particular we can construct
covariant derivatives for $(\phi^{{\bar i} {\bar j}}, \phi^{\dagger ij})$ using the starting connection $(-\nabla_i M M^{-1}, M^{\dagger -1}\bnabla_{\bar i} M^\dagger )$.}
These derivatives are also Levi-Civita covariant.

There is another useful way to write the parametrization
(\ref{cp2-13}). Towards this we first note the identities,
\beqar
{\bar D} _{\bar j} \phi^{{\bar i}{\bar j}} &=& \bnabla_{\bar j} \phi^{{\bar i} {\bar j}}
+ [ M^{\dagger -1} \bnabla_{\bar j} M^\dagger, \phi^{{\bar i} {\bar j}} ]\nonumber\\
&=& M \left[ \bnabla_{\bar j} (M^{-1} \phi M)^{{\bar i} {\bar j}}  + [ H^{-1}\bnabla_{\bar j} H,
(M^{-1} \phi M)^{{\bar i} {\bar j}} ] \right] M^{-1}\nonumber\\
&=& M \left( {\bar \D}_{\bar  j} (M^{-1} \phi M )^{{\bar i} {\bar j}} \right) M^{-1}
= M \left( {\bar \D}_{\bar j} \chi^{{\bar i} {\bar j}} \right) M^{-1}
\label{cp2-15}\\
D_j {\phi}^{\dagger ij} &=&M^{\dagger -1}  \left( \D_j \chi^{\dagger i j} \right) M^\dagger
\nonumber
\eeqar
where $\chi^{{\bar i} {\bar j}} = \epsilon^{{\bar i}{\bar j}} (M^{-1} \phi M )$,
${\chi}^{\dagger ij} = \epsilon^{ij} (M^{\dagger} \phi^\dagger  M^{\dagger -1})$. 
Further, $H$ is given as
$H = M^\dagger M$ and ${\bar \D}_{\bar j}$, $\D_j$ are defined
 with the connections 
$H^{-1} \bnabla_{\bar j} H$, $-\nabla_j H H^{-1}$,
\beqar
{\bar \D}_{\bar j} \Phi &=& \bnabla_{\bar j} \Phi + [ H^{-1} \bnabla_{\bar j} H, \Phi]\nonumber\\
{\D}_{j} \Phi &=& \nabla_{ j} \Phi + [ -\nabla_j H H^{-1}, \Phi]
\label{cp2-16}
\eeqar
Using these identities, we can write
(\ref{cp2-13}) as
\beqar
A_i &=& - \nabla_i M M^{-1} + M \left( g_{i {\bar i}} {\bar \D}_{\bar j} \chi^{{\bar i} {\bar j}} \right) M^{-1}  \nonumber\\
\bA_{\bar i} &=& M^{\dagger -1} \bnabla_{\bar i} M^\dagger
+ M^{\dagger -1} \left( -g_{{\bar i} i} \D_j \chi^{\dagger i j} 
\right) M^\dagger
\label{cp2-17}
\eeqar
These equations can be re-expressed as
\beqar
A_i &=& -\nabla_i M M^{-1} - M a_i M^{-1}\nonumber\\
\bA_{\bar i} &=& M^{\dagger -1} \bnabla_{\bar i} M^\dagger 
+ M^{\dagger -1} \ba_{\bar i} M^\dagger
\label{cp2-18}\\
a_i &=& - g_{i {\bar i}}{\bar \D}_{\bar j} \chi^{{\bar i} {\bar j}} , \hskip .3in
\ba_{\bar i} = - g_{{\bar i} i } \D_j {\chi}^{\dagger ij} = {a_i}^\dagger
\nonumber
\eeqar
It is easy to see that
$a_i$, $\ba_{\bar i}$ obey the following conditions:
\beq
g^{{\bar k} i } {\bar \D}_{\bar k} a_i = - {\bar \D}_{\bar i} {\bar \D}_{\bar j} \chi^{{\bar i} {\bar j}} = 0, \hskip .2in
g^{k {\bar i}} \D_k {\ba_{\bar i}} 
= 0
\label{cp2-19}
\eeq

The gauge-invariant degrees of freedom are now easily identified as
$H = M^\dagger M$ and 
$\chi = M^{-1} \phi \, M$,
${\chi^\dagger} = M^\dagger \phi^\dagger \, M^{\dagger -1}$.
Equivalently, they may be taken as
$H = M^\dagger M$ and $a_i$, $\ba_{\bar i}$, where
the latter are subject to the conditions
(\ref{cp2-19}).
Yet another equivalent choice would be
$\chi' =  M^\dagger \phi M^{\dagger -1}$,
$\chi'^\dagger = M^{-1} \phi^\dagger M$ and $H = M^\dagger M$.
These fields constitute the coordinates for the space of gauge-invariant
configurations, i.e., coordinates for the gauge-orbit space $\C$.

\subsection{Uniqueness of the parametrization of fields}

We now comment on the uniqueness of the parametrization of fields we have introduced. It is useful to consider the Abelian case first. The analysis based on group theory shows that the only representations of $SU(3)$ which contain a state transforming as
a vector are of the $(p,p)$-type (for which we take a derivative) and of the
$(p+3,p)$- or $(p,p+3)$-types.  This means that any vector can be parametrized as given in (\ref{cp2-10}).\footnote{Notice that this may also be viewed as a holomorphic version of
the Hodge decomposition for
one-forms in terms of an exact form, a co-exact form and a harmonic form.
 There is no ``harmonic term" for us, since the
Betti number $b_1$ of $\mathbb{CP}^2$ is zero.}

Conversely, given $A_i$, we notice that
\beq
\eta^{i {\bar i}} R_{\bar i} A_i = -\eta^{i {\bar i}} R_{\bar i} R_i f -
\e^{{\bar i} {\bar j} } R_{\bar i} R_{\bar j} \chi
= -\eta^{i {\bar i}} R_{\bar i} R_i f 
\label{resp2}
\eeq
because $[R_{\bar 1}, R_{\bar 2} ] = 0$. Since  $\eta^{i {\bar i}} R_{\bar i} R_i$
is invertible (in fact the Green's function for this will  be given later in this paper),
we can find $f$ in terms of derivatives
of $A_i$. Once we have $f$, we can rewrite (\ref{cp2-10}) as
\beq
\e^{i j} A_j = -\e^{i j} R_j f + \eta^{{\bar j} i} R_{\bar j} \chi
\label{resp3}
\eeq
which leads to 
\beq
\e^{ij} R_i A_j =  \eta^{i {\bar i}} R_i R_{\bar i} \chi
\label{resp4}
\eeq
We can now invert this to obtain $\chi$ in terms of $\e^{ij} R_i A_j $.
(The Green's function for this case, namely, with $Y = \pm 2$, is given in
\cite{KMN2}.)
Thus given $A_i$ (and its conjugate), we can determine $f$ and $\chi$
(and their conjugates). They will, of course, be nonlocal in terms
of $A$'s as expected. These arguments show the uniqueness of the parametrization for the Abelian case.

Going to the nonabelian case, we note that the term $R_i f$ is of the form 
of an infinitesimal  (complex) gauge transformation. Taking $\theta =  f$ to be Lie-algebra valued, $-R_i f = -R_i M M^{-1}$ for $M\approx 1 + \theta$.
We can ``integrate" this to a finite transformation to the form
$-R_i M \, M^{-1}$, $M = e^\theta$. The multiplication of functions on
$\mathbb{CP}^2$ with functions is still a function, so there is no
difficulty in doing this.

The remaining term in $A_i$ should be Lie-algebra-valued and transform homogeneously under gauge transformations, so we are led to the form
(\ref{cp2-13}), where we have to
(gauge) covariantize the derivative acting on $\chi$, as in (\ref{cp2-14}), or in group theory language,
${\cal R}_{\bar i} \Phi = R_{\bar i} \Phi + [ M^{\dagger -1} R_{\bar i} M, \Phi]$.
In terms of these derivatives, (\ref{cp2-13}) reads
\beqar
A_i &=& - R_i M M^{-1} - \eta_{i {\bar i}} {\cal R}_{\bar j} \phi^{{\bar i} {\bar j}}
\nonumber\\
\bA_{\bar i}&=&- M^{\dagger -1} R_{\bar i} M^\dagger - \eta_{{\bar i} i} {\cal R}_j
\phi^{\dagger i j}
\label{resp5}
\eeqar
The key point is that the
covariant derivative ${\cal R}_{\bar i}$ has no curvature, i.e.,
$[ {\cal R}_{\bar i}, {\cal R}_{\bar j} ] = 0$.
Therefore, we get
\beq
\eta^{{\bar i} i} {\cal R}_{\bar i} A_i  = - \eta^{{\bar i} i} {\cal R}_{\bar i} ( R_i M M^{-1})
\label{resp6}
\eeq
We can solve this iteratively in powers of $A$ by starting with $M \approx 1 + \theta$, since
$\eta^{{\bar i} i} {R}_{\bar i} R_i$ is invertible.
Again once we have $M$ (and $M^\dagger$ as its conjugate)
we can use
\beq
\e^{ij} R_i A_j + \e^{ij} R_iM M^{-1}\, R_jM M^{-1} = 
-\e^{ij} \eta_{j {\bar j}} R_i {\cal R}_{\bar k} \phi^{{\bar j} {\bar k}}
\label{resp7}
\eeq
The leading term on the right hand side
is $-\e^{ij} \eta_{j {\bar j}} \e^{{\bar j} {\bar k}}
R_i {R}_{\bar k} \chi = \eta^{i {\bar k}} R_i R_{\bar k} \chi$
and since $\eta^{i {\bar k}} R_i R_{\bar k}$ is invertible, again, we can, at least in principle, calculate $\chi$ in terms of $A_i$ as well.
We have thus shown how we can go from $A_i$, $\bA_{\bar i}$ to
$M$, $M^\dagger$, $\chi$, $\chi^\dagger$ and vice versa, showing uniqueness of the parametrization.

There is another feature of the parametrization (\ref{cp2-13}) or
(\ref{cp2-18}) which will be important later. 
Notice that $(M, a_i, M^\dagger, \ba_{\bar i})$ and
$(M{\bar V}(\bx ), {\bar V}^{-1}(\bx) a_i {\bar V}(\bx), V(x) M^\dagger , V(x) \ba_{\bar i} V^{-1}(x))$ lead to the same gauge potentials, where
$V(x)$ is an $SL(N, \mathbb{C})$-matrix 
 with
elements which are holomorphic functions, a holomorphic matrix for short, and ${\bar V}(\bx )$ is an antiholomorphic
matrix.
On $\mathbb{CP}^2$, there are no globally defined holomorphic or
antiholomorphic functions, except for a constant. 
Thus globally, we have no such possibility of 
$M \rightarrow M {\bar V}, \, M^\dagger \rightarrow V M^\dagger$ and
there are no additional degrees of freedom which could arise from this.

(We may note however that matrices ${V}$ (or ${\bar V}$) defined as (anti)holomorphic in {\it local}
neighborhoods can be useful to write nonsingular expressions for fields,
 in the same way that gauge transformations on
intersections of coordinate patches can be used as transition functions
for gauge fields which are specified patchwise.
The use of ${V}$ (or ${\bar V}$) as transition functions does not introduce additional functional degrees of freedom; they also do not show up in
$A_i$, $\bA_{\bar i}$.
The metric and the expression for the volume element we calculate
are also insensitive to $V$ and ${\bar V}$
since our regularization preserves the correct transformation properties
under these (anti)holomorphic transformations.
A two-dimensional example of how the local use of
$V$ and ${\bar V}$ can be useful is given in
\cite{KKN}.)

\subsection{A scalar field on $\mathbb{CP}^2$}

We now consider a massless scalar field multiplet
$\Phi$ on $\mathbb{CP}^2$, with components $\Phi^\a$
which transform under gauge
transformations according to some representation of the gauge group
$SU(N)$,
\beq
\Phi^\a\rightarrow \Phi^{'\a} = U^\a_{~\beta}\, \Phi^\beta,
\label{cp2-20}
\eeq
$U^\a_{~\beta}$ being the representation matrices corresponding to $U$
in the specific representation. The corresponding covariant derivatives are $(\nabla_i + A_i )\Phi$, $(\bnabla_{\bar i} + \bA_{\bar i}) \Phi$.
Before we write down the action, it is useful to discuss the volume element
for $\mathbb{CP}^2$. Local complex coordinates $z^i$, $ \bz^{\bar i}$,
$i = 1, 2$,
can be introduced by
taking the $3\times 3$ matrix $g$ to be such that
\beq
g_{13} = {z_1 \over \sqrt{1+\bz\cdot z}}, \hskip .2in
g_{23} = {z_2 \over \sqrt{1+\bz\cdot z}}, \hskip .2in
g_{33} = { 1\over \sqrt{1+\bz\cdot z}}
\label{cp2-21}
\eeq
The metric, which is the restriction to $\mathbb{CP}^2$ of the
Cartan-Killing metric on $SU(3)$, is given by the Fubini-Study metric
\beq
ds^2 = \left[ {d\bz\cdot dz \over 1+ \bz \cdot z}
- { z\cdot d\bz \, \bz \cdot dz \over (1+ \bz \cdot z)^2}\right]
= g_{a \ba} dz^a d\bz^{\ba}
\label{cp2-22}
\eeq
We will use volume elements normalized so that the total volume is
$1$. It is then given by
\beq
d\mu  = {2\over \pi^2} {d^4 x \over (1+ \bz \cdot z)^3}
= {2\over \pi^2} (\det g) \, {d^4 x}
\label{cp2-23}
\eeq
where $z^1=x^1-ix^2,~z^2=x^3-ix^4$.
The use of this volume element is  equivalent to using the Haar measure
on $SU(3)$, again normalized to unity.
We will consider a massless scalar with an action of the form
\beqar
S_1 &=& \int d\mu\, g^{{\bar i} i} \bigl[(\nabla_{i}  + A_{i}  )\Phi\bigr]^\dagger
\bigl[(\nabla_i + A_i )\Phi\bigr]\nonumber\\
&=& \int d\mu\, g^{{\bar i} i} \bigl[\bnabla_{\bar i}  \Phi^\dagger -  \Phi^\dagger \bA_{\bar i}  \bigr]
\bigl[(\nabla_i + A_i )\Phi\bigr]
\label{cp2-24}
\eeqar
Notice that, upon carrying out an integration by parts,
the action can be written as
\beq
S_1 = \int d\mu\, g^{{\bar i} i} \,\Phi^\dagger\,
\bigl[- (\bnabla_{\bar i}  + \bA_{\bar i}  )(\nabla_i + A_i )\bigr] \,\Phi
\label{cp2-25}
\eeq
The relevant kinetic energy operator is thus
$- g^{ {\bar i} i} (\bnabla_{\bar i}  + \bA_{\bar i}  )(\nabla_i + A_i )$. One can also
consider an action of the form
\beq
S_2 = \int d\mu\, g^{i {\bar i} } \,\Phi^\dagger\,
\bigl[- (\nabla_i + A_i ) (\bnabla_{\bar i}  + \bA_{\bar i}  ) \bigr]\,\Phi
\label{cp2-26}
\eeq
This differs from the previous one by a term of the form
$\Phi^\dagger g^{i{\bar i}} F_{i {\bar i}} \Phi $ where
$F$ is the field strength for the gauge field.
Notice that either action is completely consistent with the isometries
of the space $\mathbb{CP}^2$, so there is no {\it a priori} reason to
favor one or the other, or any linear combination of the two.
These actions essentially correspond to
chiral scalars. One can also consider a nonchiral action of the
form
\beq
S_3 = {1\over 2} \int d\mu\, g^{i {\bar i} } \,\Phi^\dagger\,
\bigl[- \bigl( (\nabla_i + A_i ) (\bnabla_{\bar i}  + \bA_{\bar i}  )
+ (\bnabla_{\bar i}  + \bA_{\bar i}  ) (\nabla_i + A_i )  \bigr)\bigr]\,\Phi
\label{cp2-27}
\eeq
In this case the kinetic operator is the Laplace operator (suitably gauge-covariantized)
on the manifold $\mathbb{CP}^2$. $S_3$ is obviously 
${\half}(S_1 + S_2)$.

To briefly summarize, this section introduced
general parametrizations for fields on $\mathbb{CP}^2$. The result for
scalar fields is given in equation (\ref{cp2-7}), 
for nonabelian gauge fields in equations (\ref{cp2-13}) or (\ref{cp2-18}).
The scalar field actions are given in
(\ref{cp2-25}), (\ref{cp2-26}) and (\ref{cp2-27}). 
Our aim now is to calculate some of the quantum corrections due to
the scalar field action, say, $S_1$,
and interpret the physical implications of the results.
This is discussed in the next section.

\section{Quantum corrections due to the scalar field}

Turning to the quantum corrections, we first note that the action
$S_1$ can be written as
\beqar
S_1
&=&\int d\mu\, g^{{\bar i} i} \left[ 
\Phi^\dagger (- \bD \cdot D ) \Phi - \Phi^\dagger M^{\dagger -1}  {\ba} M^{\dagger} \cdot D \Phi + \Phi^\dagger \bD \cdot M a M^{-1} \Phi\right. \nonumber\\
&&\left. \hskip .3in+ \Phi^\dagger M^{\dagger -1}  {\ba} M^{\dagger}M a M^{-1} \Phi \right]\nonumber\\
&=& \int d\mu\, g^{{\bar i} i} \,
\Phi^\dagger (- \bD \cdot D ) \Phi + S_{\rm int}
\label{18.2}\\
S_{\rm int}&=& \int d\mu\, g^{{\bar i} i} \left[  - \Phi^\dagger M^{\dagger -1}  {\ba} M^{\dagger} \cdot D \Phi + \Phi^\dagger M a M^{-1} \cdot \bD \Phi
+ \Phi^\dagger M^{\dagger -1}  {\ba} M^{\dagger}M a M^{-1} \Phi \right]\nonumber
\eeqar
where in getting $S_{\rm int}$ we used the transformation property of $a$ in (\ref{cp2-19}), $\bD \cdot (M a M^{-1}) = M ({\bar \D} \cdot a) M^{-1} = 0$.

Denoting the ($(M, M^\dagger)$-dependent) propagator
as
\beq
\la \Phi (x) \Phi^\dagger (y) \ra = \left( {1\over (- \bD \cdot D )}\right)_{x,y}
= \G (x, y),
\label{18.3}
\eeq
the effective action resulting from integrating out the scalar fields is
$\Gamma = \Gamma_1 + \Gamma_2$, given by
\beqar
e^{- \Gamma} &=& \int [d\Phi d\Phi^\dagger] \, e^{-S_1}\nonumber\\
\Gamma_1&=&\Tr \log (- \bD \cdot D) \nonumber\\
\Gamma_2&=& \Tr \log
\left[ 1 + M^{\dagger -1} \ba M^\dagger \cdot (- D \G (x,y))
+ M a M^{-1} \cdot \bD \G(x,y) \right.\nonumber\\
&&\hskip .3in\left. + M^{\dagger -1} \ba M^\dagger \,M a M^{-1} \G(x,y)
\right]
\label{18.4}
\eeqar
The first term $\Gamma_1$ will generate terms which only depend on
$M, M^\dagger$. The second term can be expanded in powers
of $(M a M^{-1}, M^{\dagger -1} \ba M^\dagger )$; it will correspond to
one-loop diagrams with $(M a M^{-1}, M^{\dagger -1} \ba M^\dagger )$ at the vertices and with $\G$ as the propagator.
Even though we have a massless field, there will be no infrared divergences in the calculation of $\Gamma$
since $\mathbb{CP}^2$ is a compact manifold of finite volume. If we rescale the coordinates $z_i, \bz_i \rightarrow {z_i/ r}, {\bz_i / r}$ to
introduce appropriately dimensionful coordinates, $1/r$ will serve as the
infrared cutoff. In the diagrammatic expansion of $\Gamma$, the first few 
terms will, however, be potentially ultraviolet divergent. These will include
terms in $\Gamma_1$ and the expansion of $\Gamma_2$ to the
quartic order in $(M a M^{-1}, M^{\dagger -1} \ba M^\dagger )$.
Our aim is to focus on these, evaluating them in a way fully consistent with gauge invariance and all the
isometries of $\mathbb{CP}^2$. 
We will see that the first term $\Gamma_1$ generates
a WZW action for $H = M^\dagger M$, with a finite coefficient with
potential UV divergences canceling out. Such a term can have significant
implications for physics since its critical points are instantons.
There will also be other terms in $\Gamma_1$ which combine with terms from $\Gamma_2$. The leading terms of $\Gamma_2$ will be quadratic
and logarithmic UV divergences. 

While it is straightforward to use the standard diagrammatic
expansion for $\Gamma_2$, for the evaluation of $\Gamma_1$ it is easier to consider its variation
in $M, M^\dagger$. We find
\beqar
\delta \Gamma_1 
&=&\int\Tr \left[ \delta ( M^{\dagger -1} \bnabla M^\dagger ) \, \la {\hat J} \ra
+ \delta ( \nabla M M^{-1}) \, \la {\hat J}^\dagger \ra \right]
\label{18.5}\\
\la{\hat J} (x) \ra &=& -\la D \Phi (x) \Phi^\dagger (y) \ra_{y\rightarrow x}
= - D_x \G (x, y)\bigr]_{y\rightarrow x}\nonumber\\
\la {\hat J}^\dagger (x)\ra &=&- \la  \Phi (x) (D\Phi)^\dagger (y) \ra_{y\rightarrow x}
= (-\bnabla_y \G (x, y) + \G(x, y) M^{\dagger -1} \bnabla_x M^\dagger )\bigr]_{y\rightarrow x}
\label{18.6}
\eeqar
The limit $y \rightarrow x$ has to be taken with the properly regularized
version of the propagator $\G(x, y)$.
The problem is thus reduced to the evaluation of the expectation values of
the currents as shown in (\ref{18.6}).
The currents are functions of $M, M^\dagger$ and obey a very useful condition related to the complex version of gauge transformations.
The covariant derivatives for $(h M, M^\dagger h^{-1} )$, where $h$ is
a complex matrix, are given by
\beq
D\big\vert_{h M} = h \,D\, h^{-1},\hskip .3in 
\bD\big\vert_{M^\dagger h^{-1}} = h\, \bD \,h^{-1}
\label{18.7}
\eeq
As a result, we find
\beq
\la {\hat J} ( hM, M^\dagger h^{-1} ) \ra = 
h\, \la {\hat J} (M, M^\dagger) \ra\, h^{-1}
\label{18.8}
\eeq
This property will be very useful for the evaluation of
$\Gamma_1$. A relation similar to
(\ref{18.8}) is what is used for the analogous calculation
in two dimensions.
We will be using a regularized version of this equation
as discussed in section 3.2.

\subsection{The propagator for the scalar field}
 
 The free scalar field $\Phi$ has the mode expansion
 \beq
 \Phi = \sum_{p, A} C^{(p,p)}_A \, \sqrt{(p+1)^3}\, D^{(p,p)}_{A, 0} (g)
 \label{cp2-28}
 \eeq
 Here $\sqrt{(p+1)^3}\, D^{(p,p)}_{A, 0} (g)$ are the normalized eigenfunctions
 of $\eta^{{\bar i} i}R_{\bar i} R_{ i} = - g^{{\bar i} i} 
 \bnabla_{\bar i} \,\nabla_i $. (Our approach here will be somewhat similar to what was done for the case of $\mathbb{CP}^1 = S^2$
 in \cite{AN2}.)
 Notice that, since
 $\ket{0}$  is invariant under $U(2)$ transformations,
 \beqar
 \eta^{{\bar i}i} R_{\bar i} R_i D^{(p,p)}_{A, 0} (g)&=&
\bra{s, A} {\hat g}\, \eta^{ {\bar i} i}  T_{\bar i} T_i \ket{0}\nonumber\\
&=& \bra{s, A} {\hat g} (\eta^{ {\bar i}i}  T_{\bar i} T_i + T_1^2 + T_2^2 +T_3^2 + T_8^2 ) \ket{0}
= \bra{s, A} {\hat g} T_a T_a  \ket{0}\nonumber\\
&=& p (p+2) \,D^{(p,p)}_{A, 0} (g)
\label{cp2-29}
\eeqar
where we have used the fact that the quadratic Casimir operator $T_a T_a$ has
the eigenvalue $p (p+2)$ for the $(p,p)$-representations.
The propagator is thus given by
\beqar
G(g, g') &=& \sum_{p, A} {(p+1)^3 \over p (p+2)} D^{(p,p)}_{A, 0} (g)
D^{*(p,p)}_{A, 0} (g')\nonumber\\
&=&\sum_{p=1} {(p+1)^3 \over p (p+2)} D^{(p,p)}_{0,0} (g'^\dagger g )
= \sum_{p=1} {(p+1)^3 \over p (p+2)} \bra{0} g'^\dagger g \ket{0}
\label{cp2-30}
\eeqar
where we have used the group property to combine the two
eigenfunctions to obtain $D^{(p,p)}_{0,0} (g'^\dagger g )$.
The summation starts at $p =1$ because the eigenfunction for $p =0$
is a zero mode for $ R_{\bar i} R_i$ and must be excluded from the
sum. The propagator thus obeys the equation
\beq
\eta^{{\bar i}i } R_{\bar i} R_i \, G (g, g') = \sum_{p=1} (p+1)^3\bra{0} g'^\dagger g \ket{0}
= \delta (g, g') - 1
\label{cp2-31}
\eeq
where $\delta (g, g')$ is the Dirac delta function 
on $\mathbb{CP}^2$, normalized with the volume element (\ref{cp2-23}).
Explicitly, $\delta (g, g') - 1 = \sum_{p=1}^\infty (p+1)^3\, D^{(p,p)}_{0,0} (g'^\dagger g )$. Notice that a (left) translation of $g$, $g'$ by the same 
$SU(3)$ transformation $h$, i.e., $g\rightarrow h g$, $g' \rightarrow h g'$,
leaves the propagator invariant.
This is the expression of the translational and rotational invariance of $G(g, g')$.

With the parametrization of $g$ as in (\ref{cp2-21}), 
$D^{(p,p)}_{0,0} (g'^\dagger g )$ is a polynomial of degree $p$ in
$\xi = (g'^\dagger g)_{33}(g'^\dagger g)^{*33} = (1 + s)^{-1}$, where
\beqar
s &=& \sigma^2_{z,y}  = {1  \over
(g'^\dagger_y g_z)_{33} (g^\dagger_z g'_y)_{33}  } - 1
= {(1 + \bz\cdot z) (1+ \by\cdot y) \over (1+ \bz\cdot y) (1+ \by\cdot z)} - 1
\nonumber\\
&=&{ (\bz - \by)\cdot (z -y ) + \bz \cdot z \, \by \cdot y - \bz\cdot y\, \by\cdot z\over
(1+ \bz\cdot y) (1+ \by \cdot z)}
\label{cp2-31a}
\eeqar
By construction this is invariant under translations on $\mathbb{CP}^2$
since it only involves $(g'^\dagger g)_{33}$ and its conjugate.
Further it is symmetric between the two points
and vanishes when $g' = g$ and also reduces to $\bz \cdot z$ when 
$y = 0$, i.e., for $g' =1$. Therefore it is the appropriate generalization of 
$\bz\cdot z$ to the square of the distance between two points.

It is possible to write down an expression for
$D^{(p,p)}_{0,0} (g )$, but we do not display this expression here
since it is still difficult to evaluate
the sum in (\ref{cp2-30}) in a closed form, except for certain special values
of $s$. Instead, we will obtain
the propagator by solving the differential equation (\ref{cp2-31}).
This can be done by writing the operator $R_i R_{\bar i}$  in
terms of $s$.  $\mathbb{CP}^2$ is
a K\"ahler manifold with the K\"ahler potential
\beq
K = \log (1 + \bz \cdot z ),
\label{cp2-32}
\eeq
so that the metric tensor can be written as $g_{a {\bar a}} = \del_a \bdel_{\ba}
K$. The metric so-obtained is the Fubini-Study metric given in (\ref{cp2-22}).
Explicitly, this metric tensor and its inverse are
\beqar
g_{a {\bar a}} &=& \left[ {\eta_{a {\bar a}} \over (1+ \bz\cdot z )}
- \eta_{a{\bar m} } \eta_{{\bar a} m } {\bz^{\bar m} z^m \over (1+ \bz\cdot z)^2}\right]\nonumber\\
g^{a{\bar a}} &=& (1 + \bz\cdot z) \left( \eta^{a{\bar a}} + z^a \bz^{\bar a}
\right)\label{cp2-33}
\eeqar
Because of the K\"ahler property, we also have
$\del_a ( g^{a {\bar a}} \det g ) = 0 =
\del_{\bar a} (g^{{\bar a }a } \det g) $, so that the operator of interest for us
is given as
\beqar
\eta^{{\bar i}i}R_{\bar i, z} R_{i,z}  \,G(z,y) &=& - {1\over \det g} \del_{\ba} ( g^{ {\bar a} a} \det g\, \del_{ a} G)\nonumber\\
&=&  - g^{ {\bar a} a} \bdel_{\bar a}  \del_a G  \nonumber\\
&=& - (1+ z\cdot \bz) \left( \bdel\cdot \del + \bz\cdot \bdel \, z \cdot \del \right) G(s(z,y))
\nonumber\\
&=& -\left[ s (1+ s)^2  G'' + (2 +s ) (1 +s) G' \right]
\label{cp2-34}
\eeqar
where, in the fourth line, we have expressed the operator as it acts on a function of $s  = \sigma^2_{z,y}$ as in (\ref{cp2-31a}), and the prime denotes the derivative with respect to $s$.
We now consider points with nonzero separations 
$\sigma^2_{z,y}$, hence nonzero $s$, so that the $\delta$-function in
(\ref{cp2-31}) has no support. Equation (\ref{cp2-31}) then becomes
\beq
 s (1+ s)^2  W'+ (2 +s ) (1 +s) W = 1, \hskip .3in
 W = G'
 \label{cp2-35}
 \eeq
 This is a first order inhomogeneous differential equation for $W$
 and can be solved using an integrating factor. A further integration then
 leads to the following expression for $G$,
 \beq
 G = - \left( C_1 - {1\over 2} \right) {1\over s} + C_1 \log s
 - {1\over 2} \log\left( {s \over 1+s}\right) + C_0
 \label{cp2-36}
 \eeq
 where $C_0$ and $C_1$ are arbitrary constants. For short separations,
 $s \ll 1$, the first term on the right dominates. In this limit, we see from
 (\ref{cp2-34}) that $ R_{\bar i} R_i$ is well approximated by
 $-\bdel\cdot\del$. To get the $\delta$-function upon the action of
 this operator, we need
 \beq
 G \approx {1\over 2\, \vert z-y\vert^2}
 \label{cp2-37}
 \eeq
(Recall that for $\mathbb{R}^4$, $-\nabla^2 $ acting on $1/(4 \pi^2 (x-x')^2)$
leads to the $\delta$-function. Here we have $-\bdel\cdot \del = -\nabla^2/4$,
so this removes a factor of 4. Further, we need a factor $\pi^2/2$, since we
are using the
$\delta$-function which integrates to 1, with
the volume normalized to 1 rather than $\pi^2/2$. This leads to the result (\ref{cp2-37}).)
In this limit the separation $s$ approaches that of flat space, namely,
 $\vert z-y\vert^2$.
We conclude from (\ref{cp2-37}) that $C_1 = 0$.

To determine $C_0$, we notice that, since the constant mode (or the zero mode) has been subtracted out in (\ref{cp2-31}), the propagator is orthogonal
to the zero mode. Thus we have the condition
\beq
\int d\mu \, G (g) = 0
\label{cp2-37a}
\eeq
Carrying out the integration, we find
$C_0 = - {3\over 4}$. The propagator for the massless scalar field on $\mathbb{CP}^2$ is
thus obtained as
\beq
G(z,y) = {1\over 2 \, s}   - {1\over 2} \log\left( {s \over 1+s}\right)
- {3\over 4}, \hskip .2in s = \sigma^2_{z,y}
\label{cp2-38}
\eeq
More details on these calculations are given in Appendix A, where the propagator for a massless scalar on $\mathbb{CP}^k$, for any $k$, has been derived.

%Since we set $g' =1$, the result (\ref{cp2-38}) corresponds to $G(z, 0)$.
%It is straightforward to generalize this to the propagator between two arbitrary points, i.e., to $G(z, y)$, with $y \neq 0$.
%In terms of $g$, $g'$, we write
%\beq
%s \rightarrow \sigma^2_{z,y}  = {1  \over
%(g'^\dagger g)_{33} (g^\dagger g')_{33}  } - 1
%={ (\bz - \by)\cdot (z -y ) + \bz \cdot z \, \by \cdot y - \bz\cdot y \by\cdot z\over
%(1+ \bz\cdot y) (1+ \by \cdot z)}
%\label{cp2-38a}
%\eeq
%By construction this is invariant under translations on ${\mathbb{CP}^2}$
%since it only involves $(g'^\dagger g)_{33}$ and its conjugate.
%Further it is symmetric between the two points and vanishes when $g' =g$ and also reduces to $\bz\cdot z$ when $y =0$, i.e., for $g' =1$. Therefore it is the appropriate generalization of $\bz \cdot z$ to the square of the distance between two points. The formula (\ref{cp2-38})
%for $G(s)$ with $s$ given by (\ref{cp2-38a}) will be the general form
%$G(z, y)$.

It is also useful to write this in terms of homogeneous coordinates for
${\mathbb{CP}^2}$. Let $Z = (Z^1, Z^2, Z^3)$, $Y=(Y^1, Y^2, Y^3)$ be the homogeneous coordinates corresponding to the points we labeled by
$z^i$, $y^i$. Then 
\beq
s = \sigma^2_{z,y} = {{\bar Z} \cdot Z \, {\bar Y} \cdot Y \over {\bar Z} \cdot Y \, {\bar Y}\cdot Z} - 1
\equiv  \sigma^2 (Z,Y)
\label{cp2-38b}
\eeq
Notice that this is invariant under the scaling $Z \rightarrow \lambda Z$,
$Y \rightarrow \lambda' Y$, $\lambda, \, \lambda' \in {\mathbb C} - \{ 0 \}$,
so that it is defined on the projective space rather than ${\mathbb C}^3$.
Scaling out $Z^3$, $Y^3$, in a particular coordinate patch
with $Z^3, Y^3 \neq 0$, we can write
\beqar
Z &=& Z^3 (z^1, z^2, 1) = Z^3 \sqrt{1+ \bz \cdot z} ~ (g_{13}, g_{23}, g_{33})
\nonumber\\
Y &=& Y^3 (y^1, y^2, 1) = Y^3 \sqrt{1+ \by \cdot y} ~ (g'_{13}, g'_{23}, g'_{33})
\label{cp2-38c}
\eeqar
where $z^i = Z^i/Z^3$, $y^i = Y^i/Y^3$. We then see that $s$ in (\ref{cp2-38b}) reduces to $s$ in (\ref{cp2-31a}). Thus $G(s)$
in (\ref{cp2-38}) with $s$ given in homogeneous coordinates as in
(\ref{cp2-38b}), (\ref{cp2-38c}) gives a globally valid expression for the propagator
for the scalar field.

\subsection{Regularizations}

With $\sigma^2_{z,y}$ as defined in (\ref{cp2-38b}), we
have a globally valid expression for the propagator on $\mathbb{CP}^2$.
We will now use this to define an ultraviolet regularization via
point-splitting, which is fully covariant, i.e., gauge-covariant and consistent with all the isometries of $\mathbb{CP}^2$.
Towards this, consider moving from $y$ to a nearby point with
coordinates $y'$, which we express in terms of the homogeneous coordinates as
\beq
Y \rightarrow Y' = Y + \alpha \left( {W\, {\bar Y}\cdot Y \over {\bar Y}\cdot W}
- Y\right)
\label{cp2-38d}
\eeq
where $\alpha$ is a small complex number and $W$ parametrizes the
shift of coordinates.
Notice that the added term has the same scaling behavior as $Y$.
We then find
\beq
1 + \sigma^2 (Z, Y') =  {(1 + \sigma^2(Z,Y) ) (1 + \alpha {\bar\alpha} \sigma^2(Y, W)) 
\over \left[ 1 + \alpha \left( {{\bar Z}\cdot W \, {\bar Y}\cdot Y \over
{\bar Y}\cdot W  {\bar Z}\cdot Y} -1\right)\right]
\left[ 1 + {\bar\alpha} \left( {{\bar W}\cdot Z \, {\bar Y}\cdot Y \over
{\bar W}\cdot Y  {\bar Y}\cdot Z} -1\right)\right]}
\label{cp2-38e}
\eeq
The strategy is to use $G (\sigma^2(Z,Y'))$ as the regularized version
of $G(\sigma^2(Z,Y))$, where we take $Y'$ to be different from
$Y$ by a small amount proportional to $\vert \alpha\vert \sim\sqrt{\epsilon}$. $\epsilon$ will serve as the regularization parameter. Thus
\beq
G_{\rm Reg} (Z, Y) = G(\sigma^2(Z, Y'))
\label{cp2-38e2}
\eeq
where we will include an angular averaging over the displacement
due to point-splitting.

We will be calculating the effective action in a derivative expansion,
so for most of the terms, it will turn out that we can take one of the points
$z, y$ to be at the origin, by virtue of translational invariance.
A transformation which implements this will be given in Appendix B.
But for now, if we take $y =0$, i.e., $Y= (0,0,1)$, we find
\beq
1+ \sigma^2(Z, Y') = {(1+ \bz\cdot z) (1 + \alpha {\bar\alpha} \bw\cdot w)\over
(1 + \alpha \bz \cdot w) (1+ {\bar \alpha} \bw \cdot z)}
= { {\bar Z} \cdot Z \, {\bar{\tilde W}} \cdot {\tilde W} 
\over {\bar Z} \cdot {\tilde W} \,{\bar{\tilde W}}\cdot Z}
\label{cp2-38f}
\eeq
where ${\tilde W} = (\alpha W_1, \alpha W_2, W_3) = 
W_3 ( \alpha w_1, \alpha w_2, 1)$.
Further it is useful to make a change of variables from $Z$ to
$Z'$ such that
\beq
{Z' \over Z'_3 {\bar {\tilde W}}_3} = {Z \over {\bar {\tilde W}}\cdot Z}
\label{cp2-38g}
\eeq
Notice that this transformation is covariant under independent scalings of $Z, {\tilde W}, Z'$. Equation (\ref{cp2-38g}) is equivalent to
\beq
Z' = \lambda \, Z, \hskip .3in \lambda = {Z'_3 {\bar {\tilde W}}_3 \over 
{\bar {\tilde W}}\cdot Z} 
\label{cp2-38g2}
\eeq
The key point is that, because of the homogeneity property,
the Fubini-Study metric is unchanged under
the change of variables in (\ref{cp2-38g}) or (\ref{cp2-38g2}),
\beq
ds^2 ( Z' , {\bar Z}') = ds^2(Z, {\bar Z})
\label{cp2-38j}
\eeq
Correspondingly, the inverse metric, the volume measure, etc. can be 
taken to be defined by $Z'$. 
Equation (\ref{cp2-38f}) can then be written as
\beq
1+ \sigma^2(Z, Y') = (1 + \bz'\cdot z') (1 + {\bar{\tilde w}} \cdot {\tilde w})
\equiv 1+  s (1+ \epsilon) + \epsilon 
\label{cp2-38h}
\eeq
where $s = \bz'\cdot z'$, and $ {\bar{\tilde w}}\cdot{\tilde w} = \e$. Thus, in terms of the coordinates defined by
$Z'$, the effect of regularization is to replace $G(s)$ by
$G (s(1+ \e) + \e)$,
\beq
G_{\rm Reg} (s) =  {1 \over 2 (s(1+ \e)  + \epsilon )}
- {1\over 2} \log \left( { s(1+ \e)  + \epsilon \over 1+ s(1+\e)  + \e }\right) - {3\over 4}
\label{cp2-38i}
\eeq
(The proper dimensions for $\e$ and $s$ can be restored by the scaling $z^i \rightarrow z^i/r$ and ${\tilde w}^i \rightarrow {\tilde w}^i/r$.)

This procedure provides a regularization which is covariant respecting the isometries of ${\mathbb{CP}^2}$, since $s$ in (\ref{cp2-38b}) 
is an invariant quantity. Equation (\ref{cp2-38i}) may be viewed as a covariant point-splitting and provides a uniform way to carry out calculations.

We now turn to the issue of gauge invariance.
So far we have discussed the free propagator. In the presence of gauge fields, the propagator is $\G(x,y) = (- \bD \cdot D )^{-1}_{x,y}$.
For most of the calculations we do, this will be expanded in powers of the gauge field as
\beq
\G(x, y) = G(x, y) + \int_{y_1} G(x, y_1) {\mathbb V}_{y_1} G(y_1, y) +
\int_{y_1, y_2} G(x, y_1) {\mathbb V}_{y_1} G(y_1, y_2) {\mathbb V}_{y_2}
G(y_2, y) + \cdots
\label{cp2-38k}
\eeq
where ${\mathbb V} = \bA\cdot \del + A \cdot \bdel + (\bdel\cdot A ) + \bA\cdot A$. In calculating currents such as $\la{\hat J} (x) \ra 
= - D_x \G (x, y)\bigr]_{y\rightarrow x}$ in (\ref{18.6}), we must ensure that the point-splitting is gauge-covariant as well. The point-splitting amounts to writing $G_{\rm Reg}(x, y) = G(x, y')$. Since $\G_{\rm Reg}(x, y)$ must transform as
$\G_{\rm Reg} (x, y) \rightarrow U(x)\, \G_{\rm Reg}(x, y)\, U^\dagger (y)$
under the gauge transformation $M \rightarrow U M$, $M^\dagger
\rightarrow M^\dagger U^\dagger$, we see that a gauge-invariant point-splitting is given by
\beqar
\G_{\rm Reg} (x, y) &=& \G (x, y')\, \P \exp \left( -\int_y^{y'} ( M^{\dagger -1} \bnabla M^\dagger - \nabla M M^{-1} ) \right)
\nonumber\\
&=& \left[ G_{\rm Reg} (x,y) + \int_{y_1}G(x, y_1) {\mathbb V}(y_1) G_{\rm Reg}(y_1, y) 
+ \cdots\right]\nonumber\\
&&\hskip .2in \times \P \exp \left( -\int_y^{y+ \delta y} ( M^{\dagger -1} \bnabla M^\dagger - \nabla M M^{-1} ) \right)
\label{cp2-38l}
\eeqar
Here $G_{\rm Reg}$ is as in (\ref{cp2-38e2}) and
$y' = y+\delta y$, with $\delta y^a \delta\by^{\bar a} \rightarrow \e\, \eta^{a {\bar a}}$ in taking the small $\e$ limit in a symmetric way. %\footnote{We will  be using this only for $G(z, 0)$, so that the limit applies 
%to the case when one of the coordinates corresponds to the origin
%in the local coordinates we use.
%As mentioned earlier, translational invariance of the result will ensure its applicability for
%the whole of ${\mathbb{CP}^2}$.}
Notice that, because the path-ordered exponential involves the integral of
one-forms, we can use local coordinates $y$, $y'$ in (\ref{cp2-38l}).

In principle, we can now calculate
$\Gamma$ according to
(\ref{18.4})-(\ref{18.6}), using the expression given above.
 But before doing that, we discuss some issues
regarding
the infrared side of calculations with 
(\ref{cp2-38e2}).
As mentioned earlier, on $\mathbb{CP}^2$, we do not expect infrared divergences.
Nevertheless, there is a subtlety we need to address.
Here we will consider only the first few terms in the expansion of
$\Gamma$, which are potentially
ultraviolet divergent, to understand the nature of counterterms which might be needed. The key point is that we cannot carry out an exact calculation
of all of the one-loop contributions. We can evaluate the first few terms
in a diagrammatic expansion (and the WZW term which is rather special).
So we need some control over the diagrams with higher numbers of
vertices.
Thus we need to develop an expansion scheme where the diagrams with more and more vertices are parametrically smaller. 
To see how this can be implemented,
a comparison with flat space is useful.
Basically, we are saying that the diagrammatic
expansion of $\Gamma$ will contain two types of terms.
The first few diagrams which are potentially ultraviolet divergent,
do not have infrared divergences even in flat space.
If we evaluate them in flat space with an infrared cutoff, they will be insensitive to this, or at worst, have a marginal (logarithmic) dependence.
The remaining terms in $\Gamma$, corresponding to higher numbers of vertices will be infrared divergent in flat space.
Such contributions, if we evaluate them with an
infrared cutoff $\lambda$, will be proportional to inverse powers of
$\lambda$. 
We can use an analogous procedure for $\mathbb{CP}^2$,
evaluating the corresponding diagrams with
an infrared cutoff $\lambda$. Since at short distances, the propagator
on $\mathbb{CP}^2$ approaches the flat space version, these will carry inverse powers of
$\lambda$ as well. The relevant parameter will then be $\lambda r^2 $, where $r$ is the $\mathbb{CP}^2$ radius, and
for $\lambda r^2 \gg 1$, these terms are parametrically small.  As we shall see in the next section the dominant term for $\lambda r^2 \gg 1$ is a WZW term, which is also UV finite.
The other dominant contributions are from the
potentially ultraviolet divergent terms.
The calculation of the effective action along these lines is very much in the spirit of Wilsonian renormalization.

The infrared cutoff can be included by using a simple integral representation for the propagator. We write
%Within the scheme outlined above, it is straightforward to set up the calculation. The regularized version of the free scalar propagator, with an infrared cut-off $\lambda$, has been evaluated in Appendix B, and is given by
\beq
G_{\rm Reg} (x, y) = {1\over r^2} \int_{\lambda r^2}^{\infty} dt\, \left[e^{- t \sigma^2_{x,y'}/r^2 }\left(
{1\over 2} + { 1 \over 2\, t} 
\bigl( 1- e^  {- t} \bigr)\right)   - {3\over 4} e^{- t}\right]
\label{cp2-53}
\eeq
We have introduced $r^2$ via the scaling of coordinates. The infrared
cutoff $\lambda$ appears as the lower limit of the integration over $t$.
When $\lambda$ is set to zero, we clearly reproduce
(\ref{cp2-38e2}). This result, combined with (\ref{cp2-38k}), can be
used for calculating the effective action.

\subsection{The WZW action}

We now discuss the explicit calculations, starting with the evaluation of
$\Gamma_1$. This requires, according to (\ref{18.5})-(\ref{18.6}), the
current $\la {\hat J}(x)\ra$. Notice that according to (\ref{18.8}) we can evaluate this by choosing $h = M^\dagger$, so that
\beq
\la {\hat J} (M, M^\dagger )\ra = M^{\dagger -1} \, \la {\hat J} (H, 1) \ra\,M^\dagger
\label{WZWa}
\eeq
For $\la {\hat J} (H, 1) \ra$, the relevant propagator (obtained by
$M \rightarrow H, ~M^\dagger \rightarrow 1$) is
\beqar
\G_{\rm Reg}(x, y) &=& \bra{x} {1\over (- \bnabla \cdot \D)}\ket{y'}
 \, \P \exp\left( \int_y^{y'} \nabla H H^{-1} \right)
\label{WZWb}
\eeqar
The current is then $\la {\hat J}(H, 1)\ra = - \D_x \G_{\rm Reg}(x, y)$ 
with $y \rightarrow x$. We can expand the expression (\ref{WZWb})
in powers of $\nabla H H^{-1}$. This leads to the result
\beqar
\la {\hat J}_a (H, 1)\ra &=& - {\pi \over 2} C\, \nabla_a H \, H^{-1}
~+ \cdots
\label{WZWc}\\
C&=&  {1\over \pi r^2} \left[ 1 - \log 2 + {3\over 2} e^{-\lambda r^2}
+ {\lambda r^2 \over 4} \right]\nonumber\\
&&\hskip .1in +{1\over \pi r^2}\left[ \left( E_1 (\lambda r^2) - E_1 (2 \lambda r^2)\right)
- {1\over 2} e^{-\lambda r^2} \left( 1- e^{-\lambda r^2}\right)\right]\nonumber\\
&&\hskip .1in +{1\over \pi r^2} \left[ {( 1- e^{-\lambda r^2})^2 \over 4 \lambda r^2}
+ \lambda r^2 ( e^{\lambda r^2} -1) E_1 (2\lambda r^2)
\right]
\label{WZWd}
\eeqar
Here $E_1$ denotes the exponential integral
\beq
E_1 (w) = \int_1^\infty {dt \over t} e^{-w t} =
e^{-w} \int_0^\infty dt {e^{- t}\over w+ t}
\label{WZW5}
\eeq
(The details of this calculation are given in Appendix B. There are additional terms which involve more powers of gradients of $H$ as indicated by
the ellipsis. Some of these terms will contribute to the $\log \e$-terms, see below.)
Going back to (\ref{18.5}), we can now write the variation of
$\Gamma_1$ (with respect to $M^\dagger$) as
\beqar
\delta \Gamma_1 &=& \int g^{\ba a}\Tr \left[ \delta (M^{\dagger -1} \bnabla_{\ba} M^\dagger )
M^{\dagger -1} \left( - {\pi \over 2} C\, \nabla_a H H^{-1} \right) M^\dagger\right] + \cdots
\nonumber\\
&=& - {\pi \over 2} C \int g^{\ba a} \Tr \left[ \bnabla_{\ba} (\delta M^\dagger M^{\dagger -1} )
\, \nabla_a H H^{-1} \right] +\cdots
\label{WZW5a}
\eeqar

We can now identify the part of $\Gamma_1$ corresponding to
(\ref{WZW5a}). The four-dimensional
WZW action is given by \cite{{Don},{NS}}
\beqar
S_{\rm wzw} (H) &=& {1\over 2 \pi} \int {\pi^2 \over 2} d\mu ~ g^{a \ba}\Tr ( \nabla_a H \, \bnabla_{\bar a} H^{-1} )
- {i \over 24 \pi} \int \omega\wedge \Tr (H^{-1} d H )^3 \nonumber\\
&=&{\pi\over 4} \int d\mu ~ g^{a \ba}\Tr ( \nabla_a H \, \bnabla_{\bar a} H^{-1} )
- {i \over 24 \pi} \int \omega\wedge \Tr (H^{-1} d H )^3
\label{WZW1}
\eeqar
where $\omega$ is the K\"ahler two-form on $\mathbb{CP}^2$ 
given in local coordinates as
\beqar
\omega &=&{i }\, g_{a \ba}\, dz^a \, d\bz^{\ba} 
\label{WZW2}
\eeqar
with $g_{a\ba}$ given by the Fubini-Study metric
(\ref{cp2-22}).  The last term in (\ref{WZW1}) is, as usual, over a five-manifold
which has $\mathbb{CP}^2$ as the boundary.
(The extra factor of $\pi^2/2$ in (\ref{WZW1}) compared to the standard normalizations used for this action is due to the fact that we normalized the volume to 1. Also, we use a slightly different convention for
the normalization of $\omega$, compared to \cite{nair}.)
It is easily verified by  direct computation that $S_{\rm wzw}$ obeys the 4d-version of the Polyakov-Wiegmann identity \cite{PW}, namely,
\beq
S_{\rm wzw} (N H) = S_{\rm wzw} (N) + S_{\rm wzw} (H) 
- {\pi \over 2} \int g^{\ba a}\Tr \left[ (N^{-1} \bnabla_{\ba} N ) \nabla_a H H^{-1} \right]
\label{WZW2a}
\eeq
Introducing a left variation of $M^\dagger$ by
$N \approx 1 + \delta M^\dagger \, M^{\dagger -1}$,
we find
\beq
\delta S_{\rm wzw} (H) = - {\pi\over 2} \int d\mu\, g^{a \ba} \Tr \left[ \bnabla_{\bar a} (\delta M^\dagger M^{\dagger -1})
\nabla_a H H^{-1} \right]
\label{WZW3}
\eeq
Comparing with (\ref{WZW5a}), we see that we can write
\beq
\Gamma_{1} = C \, S_{\rm wzw} (H) ~+ \cdots
\label{WZW4}
\eeq
The coefficient $C$ is as given in (\ref{WZWd}) and is finite.
It is useful to simplify it 
for limiting values of $\lambda r^2$.
For small $\lambda r^2$, we can use the expansion
$E_1 (w) \approx -\gamma - \log w + \cdots$ to obtain
\beq
C \approx  {1\over \pi r^2} \left[ {5 \over 2} - {5 \lambda r^2\over 2 }
\right] , \hskip .2in \lambda r^2 \ll 1
\label{WZW6}
\eeq
This shows that, as $\lambda \rightarrow 0$, we still get a finite
result with no infrared divergence, consistent with the expectations for
a compact space of finite volume.
For $\lambda r^2 \gg 1$, which is the  case of interest to us in view of the discussion at the end of subsection 3.2,
\beq 
C \approx {1 \over \pi r^2} \left[ 1 - \log 2
+ {\lambda r^2 \over 4} \right], \hskip .2in \lambda r^2 \gg 1
\label{WZW7}
\eeq

A number of remarks are in order at this point.
First of all, we have only evaluated $\la {\hat J} (M, M^\dagger )\ra$.
The term in $\delta \Gamma_1$, eq. (\ref{18.5}), where we vary $M$ can be obtained
via hermitian conjugation of the term resulting from $\la {\hat J} (M, M^\dagger )\ra$.
We can then verify, via the identity (\ref{WZW2a}), that
the WZW term of $\Gamma_1$ is consistent with the variation
of $M$ as well.

A second point is the following. The result (\ref{WZW4}) was obtained
by choosing $h = M^\dagger$ and using (\ref{WZWa}) for the current.
We can then ask the question whether we obtain the
same result if we use the identity
(\ref{18.8}) with $h = M^{-1}$, thus setting $M \rightarrow 1$,
$M^\dagger \rightarrow H$. In this case $D \rightarrow \nabla$,
and the relevant propagator is $(-{\bar\D} \cdot \nabla )^{-1}$.
In this case, it is not possible to obtain the result
(\ref{WZWc}) in any expansion of $(-{\bar\D} \cdot \nabla )^{-1}$ in powers
of $H^{-1} \bnabla H$ to any finite order. A resummation of an infinite series
of terms is necessary. With the resummation, we do get the same result.
The situation is similar to what happens in two dimensions.
(A more detailed explanation is given in Appendix B.)

Finally, we note that the leading term of $S_{\rm wzw}$ is negative definite.
Thus, in $e^{-\Gamma}$, which is to be used for the subsequent integration over
the gauge fields, it has the ``wrong" sign, leading to divergent
functional integrals. What this means of course is that higher terms 
in gradient of $H$ are not negligible in regimes where
integration over $H$ starts diverging (which can happen when the gradients of $H$ become large). Also, there is a similar WZW term which arises in the
calculation of the functional 
measure for the gauge fields, which is analyzed in the follow up paper
\cite{KMN2}.
It turns out that the coefficient of the combined WZW terms has 
the appropriate sign to ensure
convergence, at least for some number of chiral scalar fields
of low dimensional representations.

\subsection{The mass term}

We now turn to terms in $\Gamma_2$, eq.(\ref{18.4}). First, we notice that similarly as for $\la {\hat J}(x)\ra$ 
we can factorize out $M^{\dagger}$ and $M^{\dagger -1}$ in the trace and
send $M \rightarrow H, ~M^\dagger \rightarrow 1$. This gives us
\beq
\Gamma_2 = \Tr \log \left[ 1 + \left( -\ba \cdot \D + H a H^{-1} \cdot \bnabla +\ba H a H^{-1} \right) \G (x,y) \right]
\label{mass1}
\eeq
where $\G = 1 / (- \bnabla \cdot \D) $. The UV divergent terms can then be calculated by 
first expanding $\Gamma_2$ and then further using the expansion of
$\G$ in terms of $\nabla H H^{-1}$.
The first set of terms will have one power of $\ba$ or
$ H a H^{-1}$. Notice that the coefficient of 
$\ba$ in (\ref{mass1}) is $(-\D \G)$ which is the current we have already discussed.
We may therefore expect a {\it finite} term of the form $\Tr [ \ba \nabla H H^{-1}]$. 
Unlike the case for $S_{\rm wzw}(H)$, this term is not invariant under
$M^\dagger \rightarrow V M^\dagger$, so it is sensitive to the ambiguity of
how $M^\dagger $ is defined.
Recall that for the contribution to $\Gamma_1$, the terms with
higher powers of $\nabla H H^{-1}$, or higher number of derivatives,
do not contribute to the leading term with the minimal number of derivatives,
i.e., $S_{\rm wzw}(H)$. However, for $\ba$, the situation is less clear, since 
we have a tensor $\chi^{\dagger i j}$. The commutator of derivatives on this
gives a term with no derivatives, albeit at the cost of a power of $1/r^2$ due to the curvature.
Presumably some combination of such terms will combine with
$\Tr [ \ba \nabla H H^{-1}]$ to produce a result insensitive to the
ambiguity $M^\dagger \rightarrow V M^\dagger$.
So calculating finite terms is rather involved requiring
the careful accounting of powers of
$1/r^2$. We do not carry this out here. Instead we will focus on the
potentially ultraviolet divergent terms.
(The only finite term with significant physical implications is 
$S_{\rm wzw}(H)$, which we have already discussed.)

The next set of terms will be of the quadratic order. It is straightforward to
work this out as
\beq
\Gamma =  {1 \over 4 \e} \int d\mu ~ g^{a \bar a} \Tr ( \ba_{\bar a} H a_a H^{-1}  ) + \mathcal{O}(\log \e )
\label{massterm}
\eeq
The leading order ultraviolet divergent term is thus a mass term for the fields $a$ and $\ba$. 

It is useful to contrast this with the situation in flat space.
Consider a scalar field $\Phi$ (in flat space) coupled to $A_\mu$. 
For the sake of the argument, we will take the field
$\Phi$ to be massive to avoid any issues of infrared divergences.
Then the quantum corrections due to $\Phi$ can also, naively, lead to a mass term for the gauge field, namely, a term of the form
$\int d^4x\,A^2 $, due to vacuum polarization effects. However, usually
we reject such a term by requiring that any term we generate via quantum corrections
should be gauge-invariant and preserve the isometries of the underlying space. In flat space, the latter condition is equivalent to requiring invariance under Poincar\'e symmetry, or the corresponding Euclidean symmetry
of 4d rotations and translations.
The mass term $\int d^4x\, A^2$ does not pass this test and hence can be avoided in any regulator (such as dimensional regularization)
which preserves the required invariances.
Notice also that since $\Phi$ has a mass, we can expand diagrams with higher numbers of vertices in powers of the inverse mass and 
the terms so generated will be local. As a result, a nonlocal
mass term of the form
\beq
\Gamma_{\rm mass} \sim \int d^4x\, \Tr \left[ F^{\mu\nu} \left({1\over - D_\a D^\a }\right)
F_{\mu\nu} \right]
\label{mass2}
\eeq
which we may think of as $\int d^4x\, A^2$ completed by an infinite series of nonlocal terms to obtain the required invariance, is also not possible.

However, if
we relax the invariance conditions, the Ward-Takahashi identities for the gauge symmetry do allow for mass terms. A classic well-known case is at finite temperature. If we use the Matsubara formalism, the relevant spacetime is
$\mathbb{R}^3 \times S^1$, which has less isometries than
$\mathbb{R}^4$. In this case, we get gauge-invariant screening masses for gauge fields,
compatible with the Ward-Takahashi identities. The situation with the present case
of $\mathbb{CP}^2$ is similar.
{\it The mass term on $\mathbb{CP}^2$
given in (\ref{massterm}) is gauge-invariant and is fully consistent with
the isometries of the underlying space.
Thus there is no a priori reason to reject it}.
The divergence also implies that it is a short distance effect and
not eliminated at large values of $r$.
So the correct way to handle this is to define a renormalized theory where such a term has a coefficient renormalized to a finite value.

\subsection{The log-divergent terms}

Calculating further terms in the expansions of $\Gamma_1$ and $\Gamma_2$ we find the following logarithmically divergent terms
\beqar
\Gamma_{\rm log \e} &=& {\rm log\, \e \over 24} \int  \Tr \left[ \left( g^{a \ba} \left(\bnabla_{\ba} (\nabla_a H H^{-1}) + [\ba_{\ba}, H a_a H^{-1}]\right) \right)^2 \right. \nonumber\\
&& \hskip .8in  - 2g^{a \ba} g^{b \bar b} \left( \bnabla_{\ba} (\nabla_{b} H H^{-1}) [ \ba_{\bar b}, H a_a H^{-1}] + \bnabla_{\ba} \ba_{\bar b} \D_a (H a_b H^{-1}) \right) \nonumber\\
&& \hskip .8in  - 2g^{a \ba} g^{b \bar b} \left( \D_a (H a_b H^{-1}) [\ba_{\ba}, \ba_{\bar b} ] - \bnabla_{\ba} \ba_{\bar b} [ H a_a H^{-1}, H a_b H^{-1}] \right) \nonumber\\
&& \hskip .8in \left. + g^{a \ba} g^{b \bar b} [H a_a H^{-1}, H a_b H^{-1}][\ba_{\ba}, \ba_{\bar b}] \right]
\label{log1}
\eeqar
The term which is independent of $a$, $\ba$ is from $\Gamma_1$
due to the terms in $\la {\hat J}\ra$  with higher powers of $\nabla H H^{-1}$
and derivatives.

Even though (\ref{log1}) is a rather complicated looking expression, it 
simplifies neatly when written in terms of the field strength tensors.
The calculation is straightforward and
$\Gamma_{\rm log \e}$ reduces to the covariant form
\beqar
\Gamma_{\rm log \e} &=& {\rm log\, \e \over 24} \int \Tr ~ g^{a \ba} g^{b \bar b} \left[ 2 F_{a b} F_{\ba \bar b} - F_{a \bar b} F_{\ba b} \right]  \nonumber\\
&=& {\rm log \, \e \over 384} \int \Tr ~ g^{\mu \lambda} g^{\nu \delta} F_{\mu\nu} F_{\lambda \delta} + {\rm log \, \e \over 16} \int \Tr ~ (g^{a \ba} F_{a \ba})^2
\label{log2}
\eeqar
where the field strength tensors are, as usual, defined as
\beqar
F_{a b} &=& [\nabla_{a} +A_a, \nabla_b + A_b]  \nonumber\\
F_{\ba \bar b} &=& [\bnabla_{\ba} + {\bar A}_{\ba}, \bnabla_{\bar b} + {\bar A}_{\bar b}]  \label{Fs}\\
F_{a \bar b} &=& [\nabla_{a} +A_a, \bnabla_{\bar b} + {\bar A}_{\bar b}]  \nonumber
\eeqar
The details of the calculations, of both (\ref{massterm}) and
(\ref{log1}), are given in Appendix C.
Notice that the first term in $\Gamma_{\rm log \e}$, in the second line of
 (\ref{log2}), is proportional to the familiar
action for gauge fields. The second term is allowed for a complex
manifold such as $\mathbb{CP}^2$, since $g^{a \ba} F_{a \ba}$ does not have to vanish. The appearance of this term is linked to the
chiral nature of the action $S_1$ for the scalar field in (\ref{cp2-25}), with a kinetic operator $-{\bar D} \cdot D$. One can verify that a non-chiral kinetic energy term $-\half (D \cdot {\bar D} + {\bar D} \cdot D )$ does not produce the last term in (\ref{log2}). Further, writing
\beqar
-{\bar D} \cdot D &=& -{ 1 \over 2} \bigl[ (D \cdot {\bar D} + {\bar D} \cdot D ) - ( D \cdot {\bar D}-{\bar D} \cdot D) \bigr] \nonumber \\
&=& -{ 1 \over 2} \bigl[ (D \cdot {\bar D} + {\bar D} \cdot D ) -  g^{a \bar{a}} F_{a \bar{a}} \bigr]
\label{dif}
\eeqar
we can trace the origin of the last term in (\ref{log2}) to the $\half g^{\bar{a} a} F_{\bar{a} a}$-term in (\ref{dif}).

\subsection{Summary of section 3}

It will be useful to have a short summary of this rather long section.
We set up the expansion scheme for calculating the effective action
$\Gamma$ obtained by integrating out the scalar fields $\Phi$, $\Phi^\dagger$. The propagator for the scalar field and its regularized form were
given in (\ref{cp2-38}) and (\ref{cp2-53}), respectively.
The result for $\Gamma$ can be summarized as
\beqar
\Gamma &=& \int \Tr \left[ {1 \over 4 \e} ~  g^{a \bar a} ( \ba_{\bar a} H a_{ a} H^{-1} ) 
 +  {\rm log \e \over 384} ~ g^{\mu \lambda} g^{\nu \delta}  F_{\mu\nu}F_{\lambda \delta} \, + {\rm log \e \over 16} ~ (g^{a \ba} F_{a \ba})^2 \right] \nonumber \\
&& + \, C \, S_{\rm wzw}(H) \,  + \, {\rm finite~terms}
\label{gamma}
\eeqar
where the coefficient $C$ is given in (\ref{WZWd}).
These are the leading terms in the following sense. The first three terms give
the potential ultraviolet divergent terms, corresponding to a mass term and
the wave function renormalization of the gauge field.
There is one finite term, which is rather special, which we have
singled out. This is the WZW action for $H$, it is the finite term with the minimal number of derivatives on the $H$-field.
The terms which we have not calculated are finite terms with 
higher numbers of derivatives on $H$ or involving
powers of $a_a$, $\ba_{\ba}$.

\section{Summary and Physical Implications}

As mentioned in the introduction, 
the manifold $\mathbb{CP}^2$ has many
features making it attractive 
for analyzing the dynamics of gauge fields.
With this in mind, we
have worked out the parametrization of the gauge potentials
on $\mathbb{CP}^2$, very much along the lines of a similar parametrization
used in two dimensions.
This allowed for a simple separation of the gauge-invariant
degrees of freedom, making it possible to perform calculations in
a manifestly gauge-invariant way.
We have also obtained the form of the (chiral) scalar field propagator
on $\mathbb{CP}^2$ and worked out the leading terms in the
effective action obtained by integrating out the scalar fields.
The result is summarized in (\ref{gamma}).

We now turn to the physical implications of the results we have 
obtained.
We start with considerations regarding the mass term with the quadratically
divergent coefficient in (\ref{gamma}).
This term is manifestly consistent with gauge invariance and, also, it preserves
all the isometries of $\mathbb{CP}^2$.
Therefore, we have no reason to reject a possible mass term.
Further, the ultraviolet singularities in a field theory are only sensitive to
local geometry, so they are essentially the same as in flat space.
The appearance of this term with
a divergent coefficient therefore shows that it will survive to the
large $r$ limit. (The fact that we have reduced isometries 
even in the large $r$ limit is important for this, unlike the situation
 in $\mathbb{R}^4$ where a mass term can be ruled out on grounds of invariance.)
The existence of the mass term implies that we have to
include a counterterm
\beq
S_{\rm mass} = \mu^2 \int g^{a \bar a} \Tr ( \ba_{\bar a} H a_a H^{-1})
\label{phys1}
\eeq
in the action for the gauge fields. We can then use $\mu^2$ to absorb the divergence and define a renormalized mass
$\mu^2_{\rm Ren} = \mu^2 + (1/4 \e)$.
The natural question is then: What value should we assign
to $\mu^2_{\rm Ren}$? Recall that the four-dimensional
nonabelian gauge theory is not defined until we pick a
dimensionful parameter which sets the basic scale for the theory.
So one option is to regard $\mu^2_{\rm Ren}$ as providing this
dimensional transmutation. In this case, other renormalization effects will 
include this parameter as an infrared cutoff for the transverse modes.
Thus the usual dimensional parameter
$\Lambda_{\rm QCD}$ will be a function of this parameter
$\mu^2_{\rm Ren}$.
Equivalently, we may take the dimensional parameter to be the usual $\Lambda_{\rm QCD}$ defined via the
one-loop renormalization of the coupling constant and regard
$\mu^2_{\rm Ren}$ as determined by the theory in terms of
$\Lambda_{\rm QCD}$.
The full effective action by construction includes quantum effects in the sense that it
determines the quantum dynamics via
its equations of motion. These are essentially the Schwinger-Dyson equations of the theory. So we can think of $\mu^2_{\rm Ren}$
as determined via the Schwinger-Dyson equations, if we have
already chosen the dimensional parameter as $\Lambda_{\rm QCD}$.

The idea of a soft gluon mass\footnote{\, If the self-energy
$\Sigma (p)$ as a function of the momentum $p$ has the property that
$\Sigma (0) = \mu^2 \neq 0$ and $\Sigma (p) \rightarrow 0$ as 
$\vert p\vert \rightarrow \infty$, it is referred to as a ``soft" mass.}
goes back to the
1980s \cite{cornwall}, but it is only more recently that systematic attempts have been made
to develop this to the level of quantitative predictions \cite{{CPB},{ABP}}.
There has been considerable evidence based on lattice simulations,
where one sees clearly that the gluon propagator in the Landau gauge saturates
to a finite nonzero value at low momenta \cite{gmass-lat}. These lattice results
 do require an explanation. On the analytical side,
 there has been a lot of effort 
 in calculating the gluon self-energy via Schwinger-Dyson equations,
 and showing that it is nonzero at zero momentum;
for a review, see \cite{ABP}.
The propagator, by construction, is gauge-dependent, but the result 
for the mass is gauge-invariant since BRST Ward-Takahashi identities are
preserved.
Our analysis, which is manifestly gauge invariant, thus provides an understanding for the possible genesis of a mass term
as expected from the lattice data and is in conformity with
the analyses done using the Schwinger-Dyson equations.

Regarding the log-divergent terms, there is not much to say, except that
it contributes to the field (or wave function) renormalization and eventually gets
folded into the running of the coupling constant.
Turning to the WZW action, we first note that if $\mu^2_{\rm Ren} \neq 0$,
then the modes corresponding to $a_a$, $\ba_{\ba}$ are not relevant at low energies and the theory is controlled primarily by $S_{\rm wzw}(H)$.
This term comes with a finite coefficient even if we take
$\lambda \rightarrow 0$, i.e., there are no infrared divergences,
the result being
\beq
\Gamma_{\rm wzw} = {5 \over 2 \pi r^2} \, S_{\rm wzw}(H)
\label{phys2}
\eeq
The kinematic regime of interest to us is however for
$\lambda r^2 \gg 1$ with the coefficient of $S_{\rm wzw}(H)$ equal to
$C$ as in (\ref{WZW7}); in this regime, terms in the effective action
with vertices of higher mass-dimension are parametrically subdominant,
as explained at the end of section 3.2. This provides a consistent
argument for the theory being controlled  by $S_{\rm wzw}(H)$.
As we noted before, the coefficient in (\ref{phys2}) or
(\ref{WZW7}) has the ``wrong" sign,
contributing a growing exponential for the subsequent integration
over the gauge fields. But this is only for the contribution due to scalar matter
fields;
as we will see  in \cite{KMN2}, there is a WZW term which arises
in the measure for the
gauge fields as well; it is of the right sign and
convergence for the integration over the gauge fields is 
obtained, at least for some number of massless chiral scalar
fields of low dimensional representations.
In this case, the low energy dynamics will be dominated by
the critical points, i.e., solutions of the equations of motion, of 
$S_{\rm wzw}(H)$.
The critical points are antiself-dual instantons,
and related to holomorphic vector bundles. Essentially, $M$ and
$M^\dagger$ define the holomorphic frames for the bundle.
This action has a long history. Originally,
Donaldson considered this action in the context of antiself-dual
instantons \cite{Don}.
The same arises in attempting to
generalize the WZW theory to four dimensions and relating it to the K\"ahler-Chern-Simons theory \cite{NS}, similar to the
WZW-CS relation in two and three dimensions \cite{witten}. 
As shown in \cite{NS}, and elaborated in \cite{{Los}, {ueno}},
this action also leads to
a holomorphically factorized current algebra, very similar to the situation
in two dimensions. Such theories have also been found in higher dimensional quantum Hall systems
\cite{KNqhe},
and are also realized as the target space dynamics of (world-sheet) $N=2$ heterotic superstrings
\cite{OV}.

As mentioned in the introduction, the correlation functions for
gauge fields seem to be dominated by instantons
at low energies. A number of numerical simulations starting with the work of
the MIT group have shown clear evidence for this, see, for example,
\cite{schafer}.
On flat $\mathbb{R}^4$, it is difficult to isolate
a part of the effective action as pertaining just to the instantons,
so it is difficult to see how instanton dominance can emerge.
By considering a complex manifold such as $\mathbb{CP}^2$
and obtaining $S_{\rm wzw}$ by integrating out fields, we
obtain some analytical evidence pointing to an instanton liquid picture.
Further discussion of these matters will be taken up 
in the second part of this work.

\bigskip
VPN thanks Joannis Papavassiliou for a useful communication regarding the
Schwinger-Dyson approach to the gluon mass.

This work was supported in part by the U.S. National Science Foundation Grants No. PHY-2112729 and No. PHY-1915053, PSC-CUNY grants and by
the Bernard B. Levine Graduate Fellowship at the City College of
New York.

\section*{Appendix A: Gauge fields and the scalar propagator on $\mathbb{CP}^k$}
\def\theequation{A\arabic{equation}}
\setcounter{equation}{0}
In this Appendix, we consider the generalization of our parametrization of
gauge fields to complex projective spaces of arbitrary dimension,
i.e., to $\mathbb{CP}^k$. We will also discuss the propagator for a scalar field on such spaces. While we do not carry out explicit calculations
of effective action or gauge-invariant measures (for arbitrary $k$), this 
analysis does serve to illustrate
that there is a uniform way to treat all $\mathbb{CP}^k$.

Regarding the parametrization of the gauge
fields,  we can proceed in a way similar to what we did
for ${\mathbb{CP}}^2$ by noting that
${\mathbb{CP}}^k$ is the group coset space $SU(k+1)/ U(k)$.
Thus functions, vectors, etc. on this space may be realized in terms of the Wigner function
$D^{(s)}_{A,B}(g) = \la s, A \vert {\hat g} \vert s, B\ra$ which is the representative of an $SU(k+1)$ element
$g$ in a general irreducible representation. We designate the
representation by $s$, and 
$A$, $B$ label  the states within the representation.
For the defining fundamental representation, $g$ is a $(k+1)\times (k+1)$ unitary matrix of unit determinant.
The generators of the group in this representation will be denoted by
$\{t_a\}$, as we did for $SU(3)$.
The subalgebra $\underline{U(k)}$ is embedded in the standard way
in the algebra of $SU(k+1)$, as the upper left block in the fundamental
representation, while the $U(1)$ generator, which is the analog of the
hypercharge is given by
\beq
Y = \sqrt{2 k \over k+1} \, t_{k^2 + 2k}
=  {1\over k+1} \left[ \begin{matrix} {\mathbb 1} &0\\ 0&-k\\ \end{matrix} \right]
\label{C01}
\eeq
The right translation operators are defined as in
(\ref{cp2-1}), with $R_i$ and $R_{\bar i}$ given by the
coset generators.

A function on ${\mathbb{CP}}^k$ must be invariant under
$U(k) \in SU(k+1)$, so it can be expanded as
\beq
F (g) = \sum_{s, A}  C^{(s)}_A  \, \la s, A \vert\, {\hat g}\, \vert s, w\ra
\equiv \sum_{s, A}  C^{(s)}_A  \, D^{(s)}_{A,0}(g)
\label{C02}
\eeq
where $C^{(s)}_A$ are arbitrary coefficients characterizing the function and the state
$\vert s, w\ra \equiv \ket{0}$ is invariant under $U(k)$.
As in the case of $SU(3)$, we can think of the carrier space
of $SU(k+1)$ representations in the tensor notation as
\beq
T^{a_1 a_2 \cdots a_p}_{b_1 b_2 \cdots b_q} \equiv
\vert a_1, a_2, \cdots a_p ; b_1, b_2 , \cdots, b_q\ra
\label{C03}
\eeq
where each index can take values $1$ to
$k+1$. But unlike the case of $SU(3)$, in general, we do not have
symmetry under permutation of the $a$'s or the $b$'s.
To obtain a $U(k)$-invariant state within such a representation, which is to be identified as
$\vert s, w\ra$ in (\ref{C02}), we will need $p = q$;
the invariant state would then correspond to the
choice of $a_i = k+1, \, b_i = k+1$; i.e.,
\beq
\vert s, w\ra \equiv \ket{0} = \vert k+1, k+1, \cdots, k+1; k+1, k+1, k+1, \cdots, k+1\ra
\label{C04}
\eeq
This will also mean that the representations of
interest for functions on ${\mathbb{CP}^k}$ are of the
$s= (p,p)$-type, and are symmetric in all
$a$'s and symmetric in all $b$'s.

The components of the gauge potential must transform in the same way as
$R_i$ and $R_{\bar i}$. 
These operators transform as the fundamental and anti-fundamental
representations of $SU(k)$ and have
$Y = 1$ and $-1$, respectively.
Thus, the gauge potentials can be expanded as in (\ref{C02}), but 
with $\vert s, w\ra$ having $Y = \pm 1$ and transforming as 
fundamental and anti-fundamental of $SU(k)$.
Since functions are $U(k)$ invariant,
derivatives of functions will have these properties and will qualify
as components of the gauge potential. As before, these will describe the gradient part (i.e., the pure gauge part and the $H$-part) of the vector potentials.
There are other choices for $\vert s, w\ra$ as well. For example, 
 a state 
 \beq
\vert s, i\ra =  \vert k+1, k+1, \cdots, k+1; i, k+1, k+1, \cdots, k+1\ra
 \label{C05}
 \eeq
with all $a$'s and $b$'s being set to $k+1$, except
for $b_1 $, which is identified as the index $i$ taking values $1$ to $k$, satisfies the requirements, with $Y = 1$.
 If $b_1$ is symmetric with the other $b$'s, then this will be obtained by acting on
 the state given in (\ref{C04}). This is the gradient part we have already mentioned.
 But we can also have states where $b_1$ is antisymmetric with all the other
 $b$'s, which are themselves symmetric among themselves.
 Such a state cannot be written in terms of the action of $R_i$ on a highest weight state
 of the form (\ref{C04}).
We also have a corresponding state in the conjugate representation given by
 \beq
 \vert s, {\bar i} \ra = \vert i, k+1, \cdots, k+1; k+1, k+1, k+1, \cdots, k+1\ra
 \label{C06}
 \eeq
 This will have $Y = -1$.
 If we are considering an Abelian gauge potential on ${\mathbb{CP}^k}$,
 we can now write it as
 \beqar
 A_i &=& -R_i \, f + \sum_{s, A} a^s_A \, \la s, A \vert \, {\hat g} \, \vert
s, i\ra = -R_i \, f +  a_i\nonumber\\
\bar{A}_{\bar{i}} &=& -R_{\bar i } \, {\bar f} + \sum_{s, A} a^{s*}_A \, \la s, A \vert \, {\hat g} \, \vert s, {\bar i} \ra = -R_{\bar i } \, {\bar f} +  {\bar a}_{\bar i}
\label{C07}
\eeqar
where $f$ is a complex function with a mode expansion
\beq
f = \sum_{s, A} \lambda_A \, \la s, A\vert\, {\hat g}\, \vert s, w\ra
\label{C08}
\eeq
Since the
product of a $U(k)$-invariant state like
$\vert s, w\ra$ with another $U(k)$-invariant state will 
contain only $U(k)$-invariant states when it is reduced to irreducible components,
we see that products of functions also qualify as functions. Thus, for a nonabelian theory we can generalize (\ref{C07}) as
 \beqar
A_i &=& - \nabla_i M\, M^{-1}   - M\, a_i \, M^{-1}\nonumber\\
  \bA_{\bar i} &=& M^{\dagger -1} \bnabla_{\bar i } M^\dagger
  + M^{\dagger -1} {\bar a}_{\bar i} M^\dagger
\label{C09}
\eeqar
where $M$ is a complex matrix taking values in the complexified gauge group and $a_i$ and ${\bar a}_i$ are given by
\beq
a_i = \sum_{s, A} a^s_A \, \la s, A \vert \, {\hat g} \, \vert
s, i\ra
\label{C010}
\eeq
with ${\bar a}_i =  a_i^\dagger$.
(In (\ref{C09}), we have also changed from $R_i$, $R_{\bar i}$ to the gradient operators, as in (\ref{cp2-11}).)

Again, it is useful (and important) to count the number of polarizations shown in the
parametrization (\ref{C07}) or (\ref{C010}).
For a $U(1)$ gauge field on a complex $k$-dimensional space such as ${\mathbb{CP}^k}$,
we need $k$ complex components or $k$ independent functions
to begin with. In $f, \, {\bar f}$, we have one complex function.
The remaining terms in (\ref{C07}), namely $a_i$ and ${\bar a}_{\bar i}$
give $k$ complex (or $2k$ real) components.
But there is a constraint, just as in the case of $\mathbb{CP}^2$, since the state
$\vert s, i\ra $ is a highest weight state. The action of a raising operator
on it, whereby the index $i$ is changed to $k+1$, vanishes because
the index $b_1$ was taken to be antisymmetric under exchange with any of the other $b$'s. This means that we have the condition $\eta^{i{\bar i}} \bar{R}_{\bar i} \, a_i = 0$.
Thus effectively, we have $k -1$ independent (complex) functions in $a_i$,
so that with the $f, \, {\bar f}$, we have a total of
$k$ complex functions as needed.

We now turn to the derivation of the propagator for a scalar field on 
$\mathbb{CP}^k$. The mode expansion for such a field was
given in (\ref{C02}), (\ref{C04}).
The propagator can be written in terms of a mode expansion
as in (\ref{cp2-30}), with the local coordinates of
$\mathbb{CP}^2$ given in (\ref{cp2-21}).
These local coordinates are related to
the homogeneous coordinates $Z$ as in (\ref{cp2-38c}).
More generally, on ${\mathbb{CP}}^k$, the required representations 
$D^{(p,p)}_{A,0}(g)$ are
polynomials in $g_{a (k+1)}$ and the conjugate $g^{*a (k+1)}$.
This implies that $D_{0,0}^{(p,p)}(g)$ is a function of
$s= \eta_{a {\bar a}} z^a \bz^{\bar a}$ (and, likewise, $D_{0,0}^{(p,p)}(g'^\dagger g)$ is a function of $s = \sigma^2(z,y)$ as defined in (\ref{cp2-31a})).
The action of the operator $\eta^{{\bar i}i}R_{\bar i} R_{i} = - g^{{\bar i}i} \bnabla_{\bar i} \nabla_{i}$ on the $U(k)$-invariant state is given by
\beq
\eta^{{\bar i}i} R_{\bar i} R_i   \,D^{(p,p)}_{A, 0} (g) = p(p+k) D^{(p,p)}_{A, 0} (g)
\label{A2-2}
\eeq
where $p(p+k)$ is the eigenvalue of the quadratic Casimir operator for a $(p,p)$-representation of $SU(k+1)$. Hence, as for $\mathbb{CP}^2$, the eigenfunction for $p=0$ is a zero mode and
it must be excluded from the mode expansion of the propagator. Thus the propagator obeys the equation
\beq
\eta^{{\bar i}i}  R_{\bar i} R_i \, G (g, g') = \delta (g, g') - 1
\label{A2-3}
\eeq

Since $\mathbb{CP}^k$ is a K\"ahler manifold, the metric tensor and its inverse are given by equation (\ref{cp2-33}). The normalized volume element is 
\beq
d\mu = {k!\over \pi^k} {d^{2k}x \over (1+ \bz \cdot z)^{k+1}}
= {k!\over \pi^k} (\det g) \, {d^{2k}x}
\label{A2-4}
\eeq
The operator of interest acting on $G$ gives us
\beqar
\eta^{{\bar i}i} R_{\bar i} R_i  \,G &=& - g^{ {\bar a}a }  \bdel_{\bar a} \del_a\, G \nonumber\\
&=& -(1+s)\left[s(1+s) G'' + (k+s) G' \right]
\label{A2-5}
\eeqar
Following the propagator calculation for $\mathbb{CP}^2$, if we consider nonzero $s$, (\ref{A2-5}) becomes
\beq
s (1+ s)^2  W'+ (k +s ) (1 +s) W = 1, \hskip .3in
W = G'
\label{A2-6}
\eeq

Using a suitable integrating factor and performing an integration on W, we get the following equation for $G$,
\beqar
G &=& - \left( C_1 - {1 \over k} \right) \sum_{n=1}^{k-1} {1 \over n} \C_n^{k-1} {1\over s^n}
+ C_1 \log s - {1 \over k} \log\left( {s \over 1+s} \right) + C_0
\nonumber\\
\C_n^{k-1} &=& {(k-1)! \over n! (k-n-1)!}
\label{A2-7}
\eeqar
where the first term is present only for $k>1$.

We fix the constant $C_1$ by looking at the short distance expansion of the propagator. As $s \ll 1$, 
$G \rightarrow - \left( C_1 - {1 \over k} \right) {1\over k-1} {1\over s^{k-1}}$ (for $k=1$, $ (C_1 - 1 / k) \log s $). 
In this limit $R_{\bar i} R_{i}$ can be approximated by the flat space operator $-\bdel\cdot \del = -\nabla^2/4$. For $\mathbb{R}^{2k}$, 
the Green function for the operator $-\nabla^2$ is $ (k-2)! / (4 \pi^k (x-x')^{2(k-1)}) $ (for $k=1$, $- { \log(x-x')^2 \over 4 \pi}$). 
Removing a factor of 4 (since we are considering $-\nabla^2/4$) and multiplying by a factor of ${\pi^k \over k!}$ from the volume normalization,
we conclude that $G$ should have the following short distance limit:
\beqar
G &\approx& {1\over k(k-1) \, s^{k-1}}, \hskip .2in k >1 \nonumber\\
G &\approx& - \log s, \hskip .7in k=1
\label{A2-8}
\eeqar
This implies that $C_1 = 0$.

To find $C_0$ we notice that $G$ is given by an expansion of modes with the eigenfunction for $p=0$ removed. 
Hence it must be orthogonal to the $p=0$ eigenfunction, which is a constant. The propagator must then obey the equation
\beqar
0 &=& \int d\mu\, G \nonumber\\
&=& \int d\mu \left[ {1\over k} \sum_{n=1}^{k-1} {1 \over n} \C_n^{k-1} {1\over s^n} 
- {1 \over k} \log\left( {s \over 1+s} \right) \right] +\, C_0
\label{A2-9}
\eeqar
Solving the integral on the left we identify the constant $C_0$ as 
\beq
C_0 = - {1\over k} \sum_{n=1}^k {1\over n}
\label{A2-10}
\eeq
Thus, the massless scalar propagator for $\mathbb{CP}^k$ is
\beq
G = {1\over k} \sum_{n=1}^{k-1} {1\over n} \C_n^{k-1} {1\over s^n} -{1\over k} \log \left( {s \over s+1} \right) - {1\over k} \sum_{n=1}^k {1\over n}
\label{A2-11}
\eeq
In particular, for $k=2$,
\beq
G = {1\over 2s} - {1\over 2} \log \left( {s\over s+1} \right) - {3 \over 4}
\label{A2-12}
\eeq
which is the same as our result in (\ref{cp2-38}) for $\mathbb{CP}^2$.

\section*{Appendix B: Calculating the expectation value of the current $\la {\hat J} \ra$}
\def\theequation{B\arabic{equation}}
\setcounter{equation}{0}

In this Appendix we go over some of the details of the calculation of
the result (\ref{WZWc}) for $\la {\hat J} \ra$. The terms we need come from the
expansion of the propagator in (\ref{WZWb}) up to terms with one
power of $\nabla H H^{-1}$. The current is then given as
\beqar
\la {\hat J} \ra &=& - \D_x \G_{\rm Reg}(x, y)\big\vert_{y\rightarrow x} \nonumber\\
&=&{\rm Term~1} + {\rm Term~2} + {\rm Term~3} + \cdots
\nonumber\\
{\rm Term~1}&=& - \nabla_{xa} G(x,y')\,
 \P \exp\left( \int_y^{y'} \nabla H H^{-1} \right)\big\vert_{y\rightarrow x}\nonumber\\
 {\rm Term~2}&=& (\nabla_a H H^{-1})_{x} \, G (x, x') 
 \label{C1}\\
 {\rm Term~3}&=& \int_z \nabla_{xa} G(x, z)g_{z}^{b {\bar b}} (\nabla_b H H^{-1} )_{z} \bnabla_{z \bar b} G (z, x')\nonumber
 \eeqar
The primed homogenous coordinate is as in (52),
\beq
X' = X + \alpha \left( {W \bar{X}\cdot X \over \bar{X}\cdot W } - X \right)
\label{C2}
\eeq
For each term we do an angular average over $\alpha$ and $W$ with the conditions that $\alpha \bar{\alpha} = \epsilon$ and $\sigma^2 \left(x,w\right) = 1$.

For ${\rm Term~1}$ in (\ref{C1}), on averaging over $\alpha$ and $w$, we get
\beq
{\rm Term~1}= - (\nabla_{b} H H^{-1})_{x} \int_{\alpha} \delta(\alpha \bar{\alpha} - \epsilon) \int_{w} \delta(\sigma^2(x,w) -1) \, \nabla_{ax} G(x,y') \big\vert_{y\rightarrow x} (x'-x)^b
\label{C3}
\eeq

We can then make a coordinate transformation $w \rightarrow w'$ such that \beq
w^a = x^a + (e^{-1}_x)^a_b { w'^b \over 1- \bar{x}\cdot w'},
\label{C3a}
\eeq
where $e^{-1}_x$ are the (inverse) frame fields at $x$ as given in
(\ref{cp2-11a}).
This sets $w' \cdot \bar{w}' = \sigma^2(x,w)$.
In group theoretic terms we are using the translational invariance of the integral to change $g^{\dagger}_{x'}g_x$ to $g^{\dagger}_{w'}$, 
effectively setting $x \rightarrow 0$ and $w \rightarrow w'$ in the integral in (\ref{C3}). 

Using the following 
\beqar 
\sigma^2(x,x') &=& \alpha w' \cdot \bar{\alpha}\bar{ w}' \nonumber\\
\nabla_{xa} \sigma^2(x,y') \big\vert_{y\rightarrow x} &=& -(1+\alpha w' \cdot \bar{\alpha }\bar{w}') \, \eta_{a \bar{a}}\, (e_{x})_{ \bar{i}}^{\bar{a}}\, \bar{\alpha} \bar{w}'^{\bar{i}} \nonumber\\
(x'-x)^b &=&(e^{-1}_{x})^{b}_{i} {\alpha w'^{i} \over 1 - \bar{x}\cdot \alpha w'},
\label{C4}
\eeqar
equation (\ref{C3}) becomes
\beqar
{\rm Term~1}= (\nabla_{a} H H^{-1})_{x} \,{\epsilon \over 2 r^2}\, \left(1+{\epsilon\over r^2}\right) \, G'\left({\epsilon\over r^2}\right) 
\label{C5}
\eeqar
where we have included the scaling $\epsilon \rightarrow \epsilon / r^2$
and
$G'(s) = {\partial G \over \partial s}$.
The scalar propagator is given by
\beq
G(s) = {1 \over r^2} \int_{\lambda r^2}^{\infty} d\rho \left[ e^{-\rho s} \left( {1 \over 2} + {1 \over 2 \, \rho} \bigl( 1 - e^ {-\rho} \bigr) \right) - {3 \over 4} e^{- \rho} \right]
\label{C6}
\eeq
so that $\rm Term~1$ is
\beqar
{\rm Term~1}&=&
 (\nabla_{a} H H^{-1})_{x} \, \left[ - {1 \over 4 \epsilon} - {1 \over 2 r^2} \right]  
\label{C7}
\eeqar

Doing a similar coordinate transformation for $w$ in $\rm Term~2$ in (\ref{C1}), the rescaled $\rm Term~2$ becomes
\beqar
{\rm Term~2}&=&  (\nabla_{a} H H^{-1})_{x} \, G\left({\epsilon \over r^2}\right)\nonumber\\
&=&  (\nabla_{a} H H^{-1})_{x} \, \left[ {1 \over 2\epsilon} - {1 \over 2r^2} \log\left(\epsilon \over r^2 \right) - {\lambda \over 2} - {3 \over 4 r^2} e^{-\lambda r^2}\right. \nonumber\\
&&\left. - {1\over 2r^2}\left(E_1(\lambda r^2) + \gamma + \log (\lambda r^2) \right) \right] 
\label{C8}
\eeqar

Finally, for $\rm Term ~3$ in (\ref{C1}), we can do two coordinate transformations: one, as above, $w \rightarrow w'$ such that $\sigma^2(x,w) = w' \cdot \bar{w}'$, and
another for $z \rightarrow z'$ such that $\sigma^2(z, x) = z' \cdot \bar{z}'$. These transformations effectively set $x \rightarrow 0$ in the integral.
Furthermore, in $\rm Term ~3$, $\nabla_{b} H H^{-1}$ is at point $z$, but since we are focusing on terms without derivatives on $\nabla_{b} H H^{-1}$, we 
evaluate it at $x$. The term then  becomes 
\beqar
{\rm Term ~3} &=& (\nabla_{b} H H^{-1})_{x} \, \int_{\alpha} \delta(\alpha \bar{\alpha} - \epsilon) \int_{w'} \delta(\vert w'\vert^2 -1) \nonumber\\
&&\times\int d\mu (z') \, G'(\vert z'\vert^2) \,(1+\vert z'\vert^2) \, G'(\sigma^2(z',w')) \,(1+\sigma^2(z',w')) \,  (-\eta_{a \bar{a}}(e_x)^{\bar{a}}_{\bar{m}} \bar{z}'^{\bar{m}}) \nonumber\\
&& \hspace{0.1in} (e^{-1}_x)^b_m \, \left({z'^m \over 1 - \bar{x} \cdot z'} - {\alpha w'^m \over 1 - \bar{x} \cdot \alpha w'}\right) \, {(1+\vert z'\vert^2)(1-\alpha w' \cdot \bar{x}) \over (1+\alpha w' \cdot \bar{z}')(1-z' \cdot \bar{x})}
\label{C9}
\eeqar

%\beqar
%{\rm Term ~3} &=& (\nabla_{b} H H^{-1})_{x} \, \int_{\alpha} \delta(\alpha \bar{\alpha} - \epsilon) \int_{w'} \delta(w'^2 -1) \nonumber\\
%&&\int_{z'} (\nabla_{xa} G(x,z')) \big\vert_{x \rightarrow 0} \,g_{z'}^{b {\bar b}} \, 
 %\bnabla_{z' \bar b} G (z', w')
%\label{C9}
%\eeqar
where
\beq
1+\sigma^2(z',w') = {(1+z' \cdot \bar{z}')(1+ \alpha \bar{\alpha} w' \cdot \bar{w}') \over (1 + \alpha \bar{z}' \cdot w') (1 +\bar{\alpha} \bar{w}'\cdot z')} 
= {\bar{Z'}\cdot Z' \bar{\tilde{W'}}\cdot \tilde{W'} \over \bar{Z'}\cdot\tilde{W'} \bar{\tilde{W'}}\cdot Z'}
\label{C10}
\eeq
and $ \tilde{W'} = (\alpha W'_1, \alpha W'_2, W'_3) = W'_3 (\alpha w'_1, \alpha w'_2, 1)$.
It is now useful to make a final change of coordinates from variables $Z'$ to 
${\tilde Z}$ given by
\beq
{ \tilde{Z} \over {\tilde Z}_3 \bar{\tilde{W'}}_3 }= {Z' \over \bar{\tilde{W'}}\cdot Z'}
\label{C11}
\eeq
Equation (\ref{C10}) can then be written as
\beq
1 +\sigma^2(z',w') = (1+\bar{\tilde{z}}\cdot \tilde{z})(1+\alpha \bar{\alpha} w' \cdot \bar{w}') = (1+\bar{\tilde{z}}\cdot \tilde{z}) (1 + \epsilon)
\label{C12}
\eeq
upon angular averaging.
$\rm Term ~3$ then simplifies to 
\beqar
{\rm Term ~3} &=& -(\nabla_{b} H H^{-1})_{x} \int d\mu ({\tilde{z}}) \Bigl[G'(\vert\tilde{z}\vert^2) \, (1 +\epsilon) G'(\vert\tilde{z}\vert^2(1+\epsilon) +\epsilon) \nonumber\\
&&\hskip .6in (1+\vert\tilde{z}\vert^2)^3  \eta_{a \bar{a}} (e_x)^{\bar{a}}_{\bar{m}} (e^{-1}_x)^b_m \bar{\tilde{z}}^{\bar{m}} \tilde{z}^m \Bigr] \nonumber\\
&=&  -(\nabla_{a} H H^{-1})_{x} \int_{0}^{\infty} ds \, s^2 \, G'(s) \,  (1+\epsilon) \, G'(s(1+\epsilon) +\epsilon)
\label{C14}
\eeqar
where in the final line, $s = \vert \tilde{z}\vert^2$. After rescaling and carrying out the integral this term becomes
\beqar
{\rm Term ~3} &=& (\nabla_{a} H H^{-1})_{x} \left[-{1\over 4\epsilon} + {1\over 2r^2}\log\left({\epsilon \over r^2} \right) + {3 \lambda \over 8} + {1 \over 2r^2}\left(E_1(2 \lambda r^2) + \gamma + \log(2 \lambda r^2) \right) \right. \nonumber\\
&&\left.+{\lambda \over 2} (1 - e^{\lambda r^2}) E_1(2 \lambda r^2) + {1 \over 4 r^2} e^{-\lambda r^2} (1 - e^{-\lambda r^2}) - {1 \over 8 \lambda r^4}(1-e^{-\lambda r^2})^2 \right]
\label{C15}
\eeqar
Combining expressions (\ref{C7}), (\ref{C8}), (\ref{C15}), we get
\beq
\la {\hat J}_a \ra = - {\pi \over 2} C\, \nabla_a H H^{-1} + \cdots
\label{C16}
\eeq
with $C$ as given in (\ref{WZWd}).

%In arriving at (\ref{C11}), we used (\ref{WZWa}) with $h = M^\dagger$,
In arriving at (\ref{C16}), we used (\ref{WZWa}) with $h = M^\dagger$,
effectively eliminating $M^\dagger$ and replacing $M $ by $H$.
What is the result if we eliminate $M$? In this case,
the relevant propagator is $(-{\bar\D} \cdot \nabla )^{-1}$
and we do not obtain
(\ref{WZWc}) in any expansion of $(-{\bar\D} \cdot \nabla )^{-1}$ in powers
of $H^{-1} \bnabla H$ to any finite order.
A resummation is then needed
but the final result is the same.
We mentioned this point in section 3.3, but here we go over the arguments
in some more detail.

We start by regarding $\la {\hat J} \ra$ as a function of $M$ and
$M^\dagger$. Then, rather than using (\ref{18.8}), consider just setting $M^\dagger = 1$ in the expression for the current.
This leads to $\la {\hat J}  (M, 1) \ra = - (\pi/2) C\, \nabla M M^{-1}$.
This calculation is the same as in arriving at (\ref{WZWc}) except that we just have $M$ now, not $H$ as in the argument of the current.
(We can view this as giving the functional derivative of $\Gamma_1$ at the point
$(M, 1)$ in the space of configurations $(M, M^\dagger )$. One may then seek to integrate functionally.) Naturally, the result $\la {\hat J} \ra = - (\pi/2) C \nabla M M^{-1}$  is not gauge-covariant (since we set $M^\dagger =1$), but we know $\la {\hat J}\ra$ should be. Clearly, this has to be obtained by $M^\dagger$-dependent correction terms.
We may then ask: what $M^\dagger$-dependent terms can we add to
$\nabla M M^{-1}$ to make it covariant? To eliminate the inhomogeneous
term in $\nabla M M^{-1}$ from the gauge transformation $M \rightarrow U M$, we need a term of the form
$- M^{\dagger -1}\nabla M^\dagger $. This gives
$ -\nabla M M^{-1} - M^{\dagger -1} \nabla M^\dagger$ and leads to the result
(\ref{WZWc}). 
Notice that we have the {\it holomorphic} derivative of
$M^\dagger$ in this expression. Since $M^\dagger $ comes with the anti-holomorphic derivative in $\bD$, various terms must combine to produce
$M^\dagger$ from $M^{\dagger -1} \bnabla M^\dagger$ (and then the holomorphic derivative) which will require an infinite series.
Effectively, the identity (\ref{18.8}) is a way of carrying out this resummation.
Another way to see the argument for the gauge-covariant expression
for the current is the following. We consider the derivative
$\bnabla_{\ba} \la {\hat J}_a\ra$. With the result 
$\la {\hat J} (M, 1) \ra = - (\pi/2) C \nabla M M^{-1}+\cdots$, this becomes
\beq
\bnabla_{\ba} \la {\hat J}_a (M, 1) \ra =  (\pi/2) C \left[
\bnabla_{\ba} (- \nabla_a M M^{-1}) \right] + \cdots
\label{WZW7a}
\eeq
The term on the right hand side is the first term in
 the field strength for the potentials,
\beqar
\F_{{\ba} a} &=& \bnabla_{\ba} (- \nabla_a M M^{-1}) 
- \nabla_a (M^{\dagger -1} \bnabla_{\ba} M^\dagger )
+ [(M^{\dagger -1} \bnabla_{\ba} M^\dagger),  (- \nabla_a M M^{-1}) ]
\nonumber\\
&=& {\bar\D}_{\ba} (- \nabla_a M M^{-1})   - \nabla_a (M^{\dagger -1} \bnabla_{\ba} M^\dagger )
\label{WZW7b}
\eeqar
Since this is the gauge-covariant version of $\bnabla_{\ba} (- \nabla_a M M^{-1}) $ with the minimal number of derivatives, we see that the 
gauge-covariant extension of (\ref{WZW7a}) is 
\beq
{\bar\D}_{\ba} \la {\hat J}_a (M, M^\dagger )\ra =
{\pi \over 2} C\, \left[ {\bar\D}_{\ba} (- \nabla_a M M^{-1})   - \nabla_a (M^{\dagger -1} \bnabla_{\ba} M^\dagger )\right] +\cdots
\label{WZW7c}
\eeq
Notice further that we have the identity
\beq
\nabla_b (M^{\dagger -1} \bnabla_{\ba} M^\dagger )
- \bnabla_{\ba} (M^{\dagger -1} \nabla_{b} M^\dagger )
+ [ M^{\dagger -1} \nabla_{b} M^\dagger , M^{\dagger -1} \bnabla_{\ba} M^\dagger ] = 0
\label{WZW7d}
\eeq
Combining this with (\ref{WZW7c}) we get
\beq
{\bar\D}_{\ba}\left[  \la {\hat J}_a (M, M^\dagger )\ra
+ {\pi \over 2} C\, \left(\nabla_a M M^{-1} + M^{\dagger -1}
\nabla_a M^\dagger \right) \right] + \cdots = 0
\label{WZW7e}
\eeq
which has the solution
\beq
 \la {\hat J}_a (M, M^\dagger )\ra = M^{\dagger -1} \left[ 
- {\pi \over 2} C\, \nabla_a H H^{-1}\right] M^\dagger
+ M^{\dagger -1} \left[ V^{-1} \nabla_{a} V\right] M^\dagger
+ \cdots
\label{WZW7f}
\eeq
where $V$ is a holomorphic matrix.
This result leads us back to (\ref{WZWc}). The second term on the right  hand side is an ambiguity corresponding to the holomorphic ambiguity in defining $M$ and $M^\dagger$ mentioned at the end of subsection 2.3.
It can be removed by redefining $M^\dagger$; the WZW action is insensitive to this. Effectively, in solving (\ref{WZW7c}) using (\ref{WZW7d}),
we are carrying out a resummation.

The situation is exactly analogous to what happens in two dimensions.
In calculating $\Tr \log  \bD$ in two dimensions, one uses
the result
\beqar
\left( {1\over \bD }\right)_{x, y} &=& M^{\dagger -1} (x) \left[
{1\over \pi (x - y)}\right]_{x, y} M^\dagger (y)\nonumber\\
&\approx& \left[
{1\over \pi (x - y)}\right]_{x, y} \left[ 1 + (y -x) M^{\dagger -1} \del M^\dagger
+\cdots\right]
\label{WZW7g}
\eeqar
If $\bD^{-1}$ is expanded in powers of $M^{\dagger -1} \bdel M^\dagger$,
clearly one needs to resum an infinite series to get the holomorphic derivative
on the right hand side.

\section*{Appendix C: Calculating the UV divergent terms}
\def\theequation{C\arabic{equation}}
\setcounter{equation}{0}

In this Appendix we will go over some of the calculations leading to the UV divergent terms in (\ref{massterm}) and (\ref{log1}). 
First, we will find the divergent terms in $\Gamma_1$ by calculating the expectation value of the current as in the appendix above.
As we have seen in Appendix B, $\Gamma_1$ has at most log-divergent terms.
Such terms can be calculated in the
large $r^2$ limit treating the space as effectively flat.\footnote{By dimensional analysis, terms that are at most log-divergent
are of the form of monomials of fields and their
derivatives of scaling dimension 4,
integrated over all space. So we can calculate them in
the flat space limit and then promote the metric and volume element
to the curved space ones to obtain the covariant expressions.
}
Following equation (\ref{18.6}),
\beqar
\la{\hat J} (H, 1) \ra (x) &=& [- \D_x \G (x, y))\bigr]_{y\rightarrow x}
\label{D1}
\eeqar
where $\D$ has the connection $-\nabla H H^{-1} $, and $\G (x,y) = \left( - \bnabla \cdot \D \right)^{-1}_{x,y}$.

Expanding the propagator $\G$ in powers of $\nabla H H^{-1}$,
\beqar
\la{\hat J} (H, 1) \ra (x) &=& \left. - \D_x \left( G(x,y') + \int_z \, G(x, z) \mathbb{X}_z G(z,y') + \cdots \right) \,  \P \exp \left( \int_y^{y'} \nabla H H^{-1} \right) \right|_{y\rightarrow x}  \nonumber \\
&=& \la {\hat J} \ra^{(1)} + \la {\hat J} \ra^{(2)} + \cdots
\label{D2}
\eeqar
where $- \bnabla \cdot \D = - \bnabla \cdot \nabla - \mathbb{X} $, or explicitly, $\mathbb{X} = - \bnabla (\nabla H H^{-1} ) - \nabla H H^{-1} \cdot \bnabla $. 
The UV divergent terms arise from the first three terms of the expansion.

For log-divergent terms we are only interested in the flat space part of the propagator $G(x,y) ={ 1 \over 2 \vert x-y\vert^2}$.
Performing similar coordinate transformations as in Appendix B we introduce the regulator in the following way
\beqar
G_{\rm Reg}(x,y) &\rightarrow& {1 \over 2(\vert x-y\vert^2 + \epsilon)}
\label{D3}
\eeqar
Using results from Appendix B and performing calculations for the log-terms,
\beqar
\la {\hat J}_a \ra^{(1)} &=& \left( {1 \over 4 \e} - { \rm log \e \over 2 r^2} \right) \nabla_a H H^{-1} \nonumber \\
\la {\hat J}_a \ra^{(2)} &=& - \left( {1 \over 4 \e} - { \rm log \e \over 2 r^2} \right) \nabla_a H H^{-1} \nonumber \\
&& + {\rm log \e \over 12} \Bigl( -2 \nabla_a \bnabla(\nabla H H^{-1} ) + 3 \nabla_a H H^{-1} \bnabla(\nabla H H^{-1} )  + \nabla \cdot \bnabla (\nabla_a H H^{-1} ) \Bigr)
\nonumber\\
\la {\hat J}_a \ra^{(3)} &=& {\rm log \e \over 12} \Bigl( - \nabla_a H H^{-1} \bnabla(\nabla H H^{-1} ) - 2 \bnabla(\nabla H H^{-1} )\nabla_a H H^{-1} \nonumber \\
&& \hskip .4in  + g^{b \bar b} [- \nabla_b H H^{-1}, \bnabla_{\bar b} (\nabla_a H H^{-1} ) ] \Bigr)  
\label{D4}
\eeqar

Gathering these terms together we find
\beqar
\la {\hat J}_a \ra (H,1) &=& - {\rm log \e \over 12} \D_a \bnabla (\nabla H H^{-1} ) 
\label{D5}
\eeqar
Looking back at $\delta \Gamma_1 $ in equation (\ref{18.5}),
and given that $\la {\hat J} (M,M^{\dagger}) \ra= M^{\dagger -1} \la {\hat J} (H, 1) \ra M^\dagger$, 
we find that
\beqar
\delta \Gamma_1 &=& \int \Tr \left[ \delta (M^{\dagger -1} \bnabla M^{\dagger} ) \la {\hat J} (M, M^{\dagger}) \ra + h.c. \right] \nonumber \\
&=& \int\Tr \left[ \bnabla (\delta M^{\dagger} M^{\dagger -1}) \, \la {\hat J} (H,1) \ra  + h.c. \right] \nonumber \\
&=& {\rm log \e \over 12} \int\Tr \left[ \delta \left( \bnabla (\nabla H H^{-1} ) \right) \bnabla (\nabla H H^{-1} ) \right] 
\label{D6}
\eeqar
This identifies $\Gamma_1$ as
\beq
\Gamma_1 = {\rm log \e \over 24} \int ~ \Tr \, \left(\bnabla ( \nabla H H^{-1} ) \right)^2
\label{D7}
\eeq

To get the divergent terms in $\Gamma_2$ the calculation scheme is similar as for $\Gamma_1$.
For mass terms we use calculations as in Appendix B, for log-terms we simplify calculations by treating the 
space as flat. 
Starting from equation (\ref{mass1}),
\beqar
\Gamma_2 &=& \Tr \log \left[ 1 + \left(- \ba \cdot \D + H a H^{-1} \cdot \bnabla + \ba H a H^{-1} \right)_x \G(x,y) \right] \nonumber \\
&=& \Tr \log \left[1 + \mathbb{Y}_x \, \G (x,y) \right] \nonumber \\
&=& \int_{x} ~ \Tr \, \left. \mathbb{Y}_x \, \G (x,y') W(y',y) \right|_{y\rightarrow x}
 - {1 \over 2} \int_{x,z} ~ \Tr \, \mathbb{Y}_x \, \G(x, z) \, \mathbb{Y}_z \, \G (z,x') W(x',x) \, \cdots \nonumber \\
&=& \Gamma_2^{(1)} + \Gamma_2^{(2)} + \cdots
\label{D8}
\eeqar
where $\mathbb{Y} = - \ba \cdot ( \nabla - \nabla H H^{-1} ) + H a H^{-1} \cdot \bnabla  + \ba H a H^{-1} $ and 
$W(y',y) =  \P \exp \left( \int_y^{y'} \nabla H H^{-1} \right)$.

As above, we expand the propagator $\G$ in terms of $\nabla H H^{-1}$  to find the following UV
divergent terms
\beqar
\Gamma_2^{(1)} &=& \left( {1 \over 2 \e} - {\rm log \e \over 2 r^2} \right) \int ~ \Tr \, \ba H a H^{-1} + {\rm log \e} \int ~ \Tr {1\over 4} \bnabla (\nabla H H^{-1} ) \ba H a H^{-1} \nonumber \\
\Gamma_2^{(2)} &=& \left( -{1\over 4 \e} + {\rm log \e \over 2 r^2} \right) \int ~ \Tr \, \ba H a H^{-1} \nonumber \\
&& \hskip .1in + {\rm log \e} \int ~ \Tr \left[ - {1\over 6} \bnabla (\nabla H H^{-1} ) \left( \ba H a H^{-1} + {1\over 2} H a H^{-1} \ba \right)
+ {1\over 4} (\ba H a H^{-1} )^2 \right. \nonumber \\
&& \hskip 1.05in \left. - {1\over 12} g^{a \ba} g^{b \bar b} \bnabla_{\ba} \ba_{\bar b} \Bigl(
\nabla_a ( H a_b H^{-1} ) + [ - \nabla_a H H^{-1}, H a_b H^{-1} ] \Bigr) \right. \nonumber \\
&& \hskip 1.05in \left. - {1 \over 12}  g^{a \ba} g^{b \bar b} \bnabla_{\ba} (\nabla_b H H^{-1} ) [ \ba_{\bar b}, H a_a H^{-1} ] \right] \nonumber \\
\Gamma_2^{(3)} &=& {\rm log \e} \int ~\Tr \, \left[
- {1\over 4} (\ba H a H^{-1} )^2 - {1\over 4} \ba \cdot H a H^{-1} H a H^{-1} \cdot \ba \right. \nonumber \\
&& \hskip .8in \left. - {1 \over 12} g^{a \ba} g^{b \bar b} \bnabla_{\ba} \ba_{\bar b} [-H a_a H^{-1}, H a_b H^{-1} ] \right. \nonumber \\
&& \hskip .8in \left. -{1\over 12} g^{a \ba} g^{b \bar b} [\ba_{\ba}, \ba_{\bar b} ] \Bigl( \nabla_a (H a_b H^{-1} ) + [-\nabla_a H H^{-1}, H a_b H^{-1} ] \Bigr) \right] \nonumber\\
\Gamma_2^{(4)} &=& {\rm log \e} \int ~ \Tr \left[{1\over 24} (\ba H a H^{-1} )^2 + {1 \over 24} ( H a H^{-1} \ba )^2 + {1 \over 6} \ba \cdot H a H^{-1} H a H^{-1} \cdot \ba \right. \nonumber \\
&& \hskip .8in \left. + {1 \over 24} g^{a \ba} g^{ b \bar b} [\ba_{\ba}, \ba_{\bar b} ] [ H a_a H^{-1}, H a_b H^{-1} ] \right]
\label{D9}
\eeqar

Combining the four terms together we get
\beqar
\Gamma_2 &=&  {1 \over 4 \e} \int ~ \Tr \, \ba H a H^{-1} \nonumber \\
&& + {\rm log \e \over 24} \int ~ \Tr \Bigl[ 2 \bnabla (\nabla H H^{-1} ) [\ba , H a H^{-1} ] 
+ [\ba , H a H^{-1} ]^2  \nonumber \\
&& \hskip .9in - 2  g^{a \ba} g^{b \bar b} \, \Bigl( \bnabla_{\ba} ( \nabla_b H H^{-1} ) [ \ba_{\bar b}, H a_a H^{-1} ] + \bnabla_{\ba} \ba_{\bar b} \D_a (H a_b H^{-1}) \Bigr) \nonumber \\
&& \hskip .9in - 2  g^{a \ba} g^{b \bar b} \, \Bigl( \D_a (H a_b H^{-1} ) [\ba_{\ba}, \ba_{\bar b}] - \bnabla_{\ba} \ba_{\bar b} [H a_a H^{-1}, H a_b H^{-1} ] \Bigr) \nonumber \\
&& \hskip .9in  + g^{a \ba} g^{b \bar b} \, [H a_a H^{-1}, H a_b H^{-1} ] [ \ba_{\ba}, \ba_{\bar b} ] \Bigr]
\label{D10}
\eeqar

Combining $\Gamma_1$ and $\Gamma_2$ from (\ref{D7}) and (\ref{D10}) above, 
the ultraviolet divergent terms are
\beqar
\Gamma_{\rm div} &=& {1 \over 4 \e} \int ~ \Tr \, \ba H a H^{-1} \nonumber \\
&& + {\rm log \e \over 24} \int ~ \Tr \, \left[ \left( \bnabla \cdot ( \nabla H H^{-1} )  + [ \ba, H a H^{-1} ] \right)^2 \right. \nonumber \\
&& \hskip .9in - 2  g^{a \ba} g^{b \bar b} \, \Bigl( \bnabla_{\ba} ( \nabla_b H H^{-1} ) [ \ba_{\bar b}, H a_a H^{-1} ] + \bnabla_{\ba} \ba_{\bar b} \D_a (H a_b H^{-1}) \Bigr) \nonumber \\
&& \hskip .9in - 2  g^{a \ba} g^{b \bar b} \, \Bigl( \D_a (H a_b H^{-1} ) [\ba_{\ba}, \ba_{\bar b}] - \bnabla_{\ba} \ba_{\bar b} [H a_a H^{-1}, H a_b H^{-1} ] \Bigr) \nonumber \\
&& \hskip .9in \left. + g^{a \ba} g^{b \bar b} \, [H a_a H^{-1}, H a_b H^{-1} ] [ \ba_{\ba}, \ba_{\bar b} ] \right]
\label{D11}
\eeqar
which is the result in (\ref{massterm}) and (\ref{log1}).

\end{document}